\documentclass[sigconf,nonacm]{acmart}

\AtBeginDocument{%
  \providecommand\BibTeX{{%
    \normalfont B\kern-0.5em{\scshape i\kern-0.25em b}\kern-0.8em\TeX}}}

\setcopyright{none}
\copyrightyear{ }
\acmYear{ }
\acmDOI{}
\newsavebox\ltmcbox

\usepackage{lineno,hyperref}
\usepackage{longtable}
\usepackage{graphicx}
\usepackage{multirow}
\usepackage{enumitem}
\usepackage{amsfonts}

\acmBooktitle{}
\acmPrice{}
\acmISBN{}

\begin{document}

\title{Deep Reinforcement Learning-based Methods for Resource Scheduling in Cloud Computing: A Review and Future Directions}

\author{Guangyao Zhou}
\email{guangyao\_zhou@std.uestc.edu.cn}
\affiliation{%
  \institution{School of Information and Software Engineering, University of Electronic Science and Technology of China}
  \city{Cheng Du}
  \country{China}
}

\author{Wenhong Tian}
\email{tian\_wenhong@uestc.edu.cn}
\affiliation{%
  \institution{School of Information and Software Engineering, University of Electronic Science and Technology of China}
  \city{Cheng Du}
  \country{China}
}

\author{Rajkumar Buyya}
\email{rbuyya@unimelb.edu.au}
\affiliation{%
  \institution{Cloud Computing and Distributed Systems (CLOUDS) Laboratory, Department of Computing and Information Systems, The University of Melbourne}
  \city{Melbourne}
  \country{Australia}
}

\author{Ruini Xue}
\email{xueruini@uestc.edu.cn}
\affiliation{%
  \institution{School of Computer Science and Engineering, University of Electronic Science and Technology of China}
  \city{Cheng Du}
  \country{China}
} 
  \author{Liang Song}
\email{songliang@tsinghua-eiri.org}
\affiliation{%
  \institution{Sichuan Energy Internet Research Institute, Tsinghua University}
  \city{Cheng Du}
  \country{China}
}

\renewcommand{\shortauthors}{ }

\begin{abstract}
With the acceleration of the Internet in Web 2.0, Cloud computing is a new paradigm to offer dynamic, reliable and elastic computing services. Efficient scheduling of resources or optimal allocation of requests is one of the prominent issues in emerging Cloud computing. Considering the growing complexity of Cloud computing, future Cloud systems will require more effective resource management methods. In some complex scenarios with difficulties in directly evaluating the performance of scheduling solutions, classic algorithms (such as heuristics and meta-heuristics) will fail to obtain an effective scheme. Deep reinforcement learning (DRL) is a novel method to solve scheduling problems. Due to the combination of deep learning (DL) and reinforcement learning (RL), DRL has achieved considerable performance in current studies. To focus on this direction and analyze the application prospect of DRL in Cloud scheduling, we provide a comprehensive review for DRL-based methods in resource scheduling of Cloud computing. Through the theoretical formulation of scheduling and analysis of RL frameworks, we discuss the advantages of DRL-based methods in Cloud scheduling.  We also highlight different challenges and discuss the future directions existing in the DRL-based Cloud scheduling.
\end{abstract}


\maketitle

\begin{sloppypar}
\section{Introduction}\label{sec1}
%\subsection{Motivation}
Cloud computing is generally accepted as a type of distributed system linked by {a} high-speed network. It includes the applications delivered as services over the Internet, the hardware and systems software that can dynamically provide services to users \citep{b21,b80}.
As a paradigm that provides services to users in a pay-as-you-go \citep{b73} manner or pay-per-use \citep{b75}, Cloud computing has four forms:  Infrastructure as a Service (IaaS), Platform as a Service (PaaS), Software as a Service (SaaS) \citep{b21,b80,b97,b125,b128}, and a new form of serverless computing \citep{b32,b80}.

Cloud computing provisions computing resources on the basis of CPU (Central Processing Unit) \citep{b80,b169}, RAM (Random Access Memory) \citep{b103,b138}, GPU (Graphics Processing Unit) \citep{b181,b83}, Disk Capacity \citep{b80,b169} and Network Bandwidth \citep{b103,b139}.
From another perspective, ``time'' and ``space'' are also two pivotal resources of Cloud computing. Time means the whole service life cycle of the Cloud platform, and space means the real physical place to emplace physical devices.
Electrical components of Cloud computing devices are driven by electric energy and work at the time and space. They constitute the real resources assembled of Cloud computing. Therefore, real natural resources provided by Cloud computing are effective electric energy conversion per unit of space and per unit of time {(frequency)}, regarding energy, time, and space as essential resources {\citep{ln1}}.
The limited resource utilization capacity of Cloud computing will raise the cost and energy consumption of Cloud system \citep{b75,b110}. Moreover, long response time, long queuing time and high delay rate will direct the decrease in QoS (quality of service).
Consequently, how to schedule components of Cloud computing in an efficient, energy-saving, balanced method, is a critical factor, influencing the orientation of Cloud computing in the future.

\begin{figure}[ht]
        \begin{minipage}[h]{1.0\columnwidth}
            \centering
            \includegraphics[width=\columnwidth]{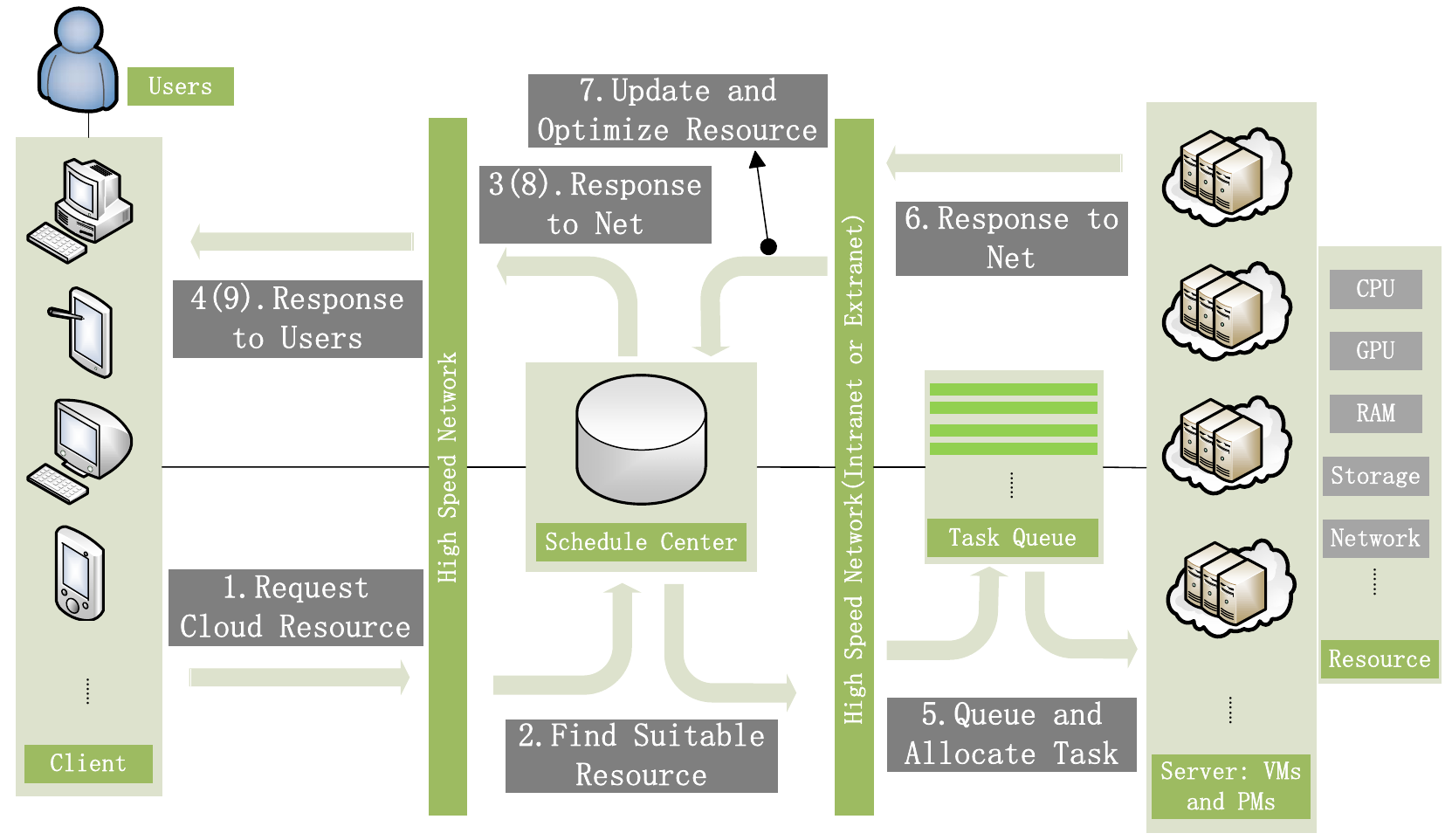}
            \caption{Resource Management and Task Allocation Process based on {Schedule Center}.}
            \label{fig0}
        \end{minipage}
\end{figure}

Cloud computing has some characteristics including the huge scale of devices, the complexity of scenarios, the unpredictability of user requests, the randomness of electronic components, and {the} uncertain temperature of various components presented in the running process. These characteristics
pose challenges to efficient and effective resource scheduling of Cloud computing \citep{b69,b71,b81}.
Currently, multi-phase approach \citep{b56,b71,b104}, virtual machine migration \citep{b76,b102,b105}, queuing model \citep{b46,b58,b73,b81}, service migration \citep{b113,b187,b97}, workload migration \citep{b67}, application migration \citep{b97,b81}, task migration \citep{b68,b76,b90} and scheduling algorithm of scheduler are current common strategies to resolve the resource management of Cloud computing. Among these, the core of the solution is still the design of the scheduler on the basis of the scheduling algorithm.  Figure \ref{fig0} shows a resource management and task allocation process with a scheduler as the core. The users operate the clients to submit task requests to the Cloud center through the high-speed networks; The Cloud center collects tasks, generates scheduling schemes leveraging scheduling algorithms, and allocates tasks to server nodes; The server nodes then provide corresponding services to users \citep{z2}.
Due to its impact on the effective operation of Cloud, scheduling algorithms of Cloud computing have attracted researchers. The scheduling problem in distributed systems is usually an NP-complete problem or an NP-hard problem without a polynomial-time algorithm unless $NP=P$ \citep{b80,b144,b139}.
Existing methods to resolve scheduling problems mainly contain six categories including Dynamic Programming, Probability algorithm, Heuristic method, Meta-Heuristic algorithm, Hybrid algorithm and Machine Learning (ML).

As classic methods (non-machine learning) are not experts in addressing the complex scheduling scenarios of Cloud computing, there are abundant discussions and research about the application of ML in Cloud scheduling such as work from Microsoft \citep{b201}, CLOUDS Laboratory of The University of Melbourne \citep{b193}, and other institutes \citep{b81,b92,b186}.
Deep reinforcement learning (DRL), belonging to ML, is a novel approach combined with the advantages of {the} deep neural network (DNN) and reinforcement learning (RL). In recent years, DRL  {has been} prevalent in solving Cloud scheduling and has been proven to occupy strong superiorities in many complex scenarios \citep{b107,b150,b153,b161,b164,b168,n12,n9}.
{There are many} surveys \citep{b42,b76,b80,b81,b92,b94,b97,b98,b99,ln6,b102,b104,b119,b182,b183,b184,b185,b186,b198,ln5,z4,ln4} that have provided detail, comprehensive and valuable reviews of various fields in Cloud computing. { Some examples related to Cloud resource optimization management are as follows. \cite{b80} reviewed the workflow scheduling in Cloud and analyzed the characteristics of its techniques by classifying them based on the objectives and execution mode.
\cite{ln4} focused on the performance interference of virtual machines and revisited interference-aware strategies for scheduling optimization as well as co-optimization-based approaches.
 \cite{b185} provided a literature survey for task scheduling strategies (mainly including some meta-heuristic algorithms-based task scheduling) and discussed the various issues related to scheduling methodologies and the limitations to overcome. 
 \cite{b99} reviewed load balancing algorithms for virtual machine placement in cloud computing, including some heuristic, meta-heuristic and hybrid algorithms related to the load balancing problems. 
Following different scheduling scenarios,  \cite{b97} presented a comprehensive survey of evolutionary approaches in Cloud resource scheduling, mainly including the genetic algorithm (GA), ant colony optimization (ACO) and particle swarm optimization (PSO).  \cite{ln5} presented a review for a taxonomy of meta-heuristic scheduling techniques in Cloud and fog, from several categories including physics-based algorithms, evolutionary algorithms, biology-based algorithms, chemistry-based algorithms, etc.}
Some existing surveys have discussed the application of ML in Cloud scheduling \citep{b198,b81,b92,b186,z4}. 
{ For example, \cite{b81}  discussed some ML methods for resource provisioning in edge-Cloud applications, mainly including the applications of DNN, support vector machines (SVM), decision trees, Bayesian networks, splines, and exponential smoothing.  \cite{b92} discussed the application of machine learning in computation and communication control in mobile edge computing, including fuzzy control model, tree-based naive Bayes, SVM, etc.  \cite{z4} presented a literature review for the application of ML methods in Cloud resource management, mainly including prediction or classification approaches such as SVM, k-nearest neighbors (KNN), DL, etc.
}
However, there is no survey specifically discussing the application of DRL in Cloud scheduling, as it is a novel direction emerging and developing in recent years.
Researchers are still exploring the application pattern of RL, especially DRL in Cloud scheduling \citep{b150,b151,b176,b177,b169,b107,b153,b83,b161,b168,b203}. Similarly, DRL (or RL) is also applied to solve scheduling problems in other field \citep{b195,b199}.

%Through a large number of search for a variety of literature databases (including dblp, Web of Science, IEEE Xplore, ScienceDirect), we observe that the number of literature on resource scheduling of Cloud computing using DRL is not large, but we also observe that the existing research using DRL has achieved considerable results.
Noting the potential application value of DRL in Cloud scheduling, we consider providing a comprehensive survey for existing research using DRL-based methods to solve Cloud scheduling. Based on the reviews and discussions, we finally target challenges and future directions using DRL to adapt to more realistic scenarios of Cloud scheduling.

The main contributions of this paper can be summarized as follows.
\begin{enumerate}[label=(\arabic*), leftmargin=*, topsep = 0em, partopsep=0em, itemsep=0em, parsep = 0em]
\item A comprehensive review and discussions of existing scheduling algorithms for Cloud computing;
\item An analysis for the frameworks of RL and DRL from the perspective of model structures;
\item A structured review and discussion of existing research using DRL in Cloud scheduling;
\item Some identified challenges and potential future directions of DRL-based methods in Cloud scheduling.
\end{enumerate}

The rest of the paper is organized as follows.
%Section \ref{sec2} briefly reviews the developing process of research of scheduling problems in Cloud computing.
%Section  \ref{sec3} establishes the mathematical models for various optimization objectives of scheduling problems in Cloud computing.
According to the classification of classic methods and machine learning methods,  {Sect.} \ref{sec4} formulates the scheduling and reviews the existing scheduling algorithms utilized in Cloud computing.
{Sect.} \ref{sec5} presents the structure analysis of RL and DRL applied in Cloud scheduling to assist better understanding of DRL (RL) methods used in the existing research.
{Sect.} \ref{sec6} provides some structured presentations of existing research using DRL methods and discusses the current situation of DRL in Cloud scheduling.
Then, {Sect.} \ref{sec7} lists challenges and potential future directions of applying DRL in Cloud scheduling.
Finally, {Sect.} \ref{sec8} concludes this paper.

\begin{table*}[ht!]
\setlength{\tabcolsep}{3pt}
\caption{A {List} of Notations with Descriptions}
\label{tableadd0}
\begin{tabular}[ht]{p{80pt}p{280pt}}

\hline
{Notations} &Descriptions \\
\hline
$M$ & Number of indivisible tasks\\
$N$ &Number of server nodes\\
$D$ & Number of dimensions for the resources in a server node\\ 
$i$ & The index of task\\
$j$ & The index of server node\\

$k$ & The index of dimension for the resources\\
$v_{ijk}$&The capacity or space or time requirement for $j$-th dimensional resource when the $i$-th task is allocated to the $j$-th service node\\
$V_i=\left\{ v_{ijk} \right\}_{N\times D}$& The parameter matrix of the $i$-th task\\
$V=\left\langle V_1, V_2,\dots V_M \right\rangle$& The set of parameters matrices of tasks \\
$l_{jk}$& The load status of the $k$-th dimensional resource in the $j$-th server node\\
$L=\left\{ l_{jk} \right\}_{N\times D}$ & The parameters matrix of server nodes\\
$X=\left\{x_{ij}\right\}_{M\times N}$ & The matrix representing the allocation of mapping ``Tasks $\to$ Resources''\\
$s_i$ & The start time of the $i$-th task\\
$e_i$ & The end time of the $i$-th task\\
$S=\left\{s_{i}\right\}_{M}$& The matrix with the start time of tasks\\
$\omega\left( X, S, V, L \right)$ &The optimization objective of scheduling mapped from $\left\langle X,S, V, L \right\rangle$\\
$Al\left(V,L,\omega\right)$ & The scheme generated by algorithm $Al$ according to the input of  $\left(V,L,\omega\right)$\\
\hline
\end{tabular}
\end{table*}

\section{Scheduling and Algorithms in Cloud}\label{sec4}
\subsection{Mathematical Formulation of Scheduling}
For {the} sake of the presentation, we list some notations with descriptions in Table \ref{tableadd0}.

In distributed systems, scheduling problems are usually NP-hard \citep{b80,b501,b44}. Some of the mainstreams in Cloud scheduling focus on objectives including minimizing energy consumption \citep{b165, b114, ln1}, minimizing makespan \citep{b162, b163, b164}, minimizing delay time (or delayed services) \citep{b93, b135, b73}, reducing response time \citep{b113, b111}, maximizing the degree of load balancing \citep{b162, b112, b84}, increasing reliability \citep{b93, b113}, increasing the utilization of resources \citep{b108, b151, b58}, maximizing the profit of providers \citep{b162, b163, b60}, maximizing task completion ratio \citep{b113, b101, b122}, minimizing Service Level Agreement (SLA) Violation \citep{b113, b108,b203}, maximizing throughput \citep{b136,b114,b117}, and multi-objectives \citep{b163,b165,b114}.

There are several different definitions of resource scheduling in some literature \citep{b76,b80,b97}. From \cite{b76}, resource scheduling can be done in two ways: first is on-demand scheduling in which the Cloud service provider provides the resources quickly to random workload, and second is long-term reservation in which large numbers of virtual machines are in ideal condition due to which under-provisioning type of problem occurs. From \cite{b80}, task scheduling is to find an optimal order of the tasks that meet the scheduling objectives. Resource scheduling is defined by \cite{b97} as to find an ``optimal" mapping ``Tasks $\to$ Resources'' to meet one or several given objectives. There are still other definitions, which focus on whether the scheduled object is a task, a workflow, or a resource. Additionally, there are also some definitions using resource scheduling as a general term for resource management which may also contain task scheduling, workflow scheduling, resource scheduling, etc. In this paper, we unify these by ``resource scheduling in Cloud computing'' or ``Cloud scheduling''. Then, a scheduling algorithm for Cloud can be defined as an algorithm with specific rules, strategies, or processes that can generate a scheduling scheme including which resources a task is assigned to (i.e., $X$), and when to start processing the task (i.e., $S$). 

Referring to existing studies of Cloud scheduling and for the sake of comprehensive discussion, we can establish a universal formulation for scheduling problems.
It can be assumed that the number of indivisible tasks is $M$, the number of server nodes is $N$, and each server node has $D$ dimensional resources (such as CPU load, GPU load, RAM, bandwidth, disk storage, etc.). Then, the $i$-th task can be represented by a parameter matrix $V_i=\left\{ v_{ijk} \right\}_{N\times D}$ where $1\le i\le M$,  $1\le j\le N$,  $1\le k\le D$, and $v_{ijk}$ indicates the capacity or space or time requirement for $j$-th dimensional resource when the $i$-th task is allocated to the $j$-th service node.  The set of parameters of tasks $\left\langle V_1, V_2,\dots V_M \right\rangle$ is set as $V=\left\{ v_{ijk} \right\}_{M\times N\times D}$. The parameters of server nodes can be set as $L=\left\{ l_{jk} \right\}_{N\times D}$, where $l_{jk}$ means the load status of the $k$-th dimensional resource in the $j$-th server node.
Using a matrix $X=\left\{x_{ij}\right\}_{M\times N}$ to represent the allocation solution of mapping ``Tasks $\to$ Resources'' and a matrix $S=\left\{s_{i}\right\}_{M}$ to represent the start time of tasks, then a scheduling scheme can be expressed by the combination of $X$ and $S$, marked as $\left\langle X, S \right\rangle$. Wherein, $x_{ij}\in\{0,1\}$ and $\sum_{j=1}^{N} x_{ij}=1$, which means the indivisible task can be allocated to only one node. $x_{ij}=1$ means the $i$-th task is allocated to the $j$-th node. Limiting $S$ can generate the execution order between tasks. For example, setting $s_{i_1}\ge e_{i_2}$ (where $e_{i}$ is the end time of the $i$-th task) equals that the $i_1$-th task must begin after the finish of the $i_2$-th task. Thus, the matrix $S$ is sufficient to include the execution order of the task. 

A optimization result of scheduling is a mapping from the solution $\left\langle X, S \right\rangle$, the parameters of tasks $V$ and server nodes $L$. Thus, the optimization objective can be set as
\begin{equation}\label{eq1}
    \min \omega=\omega\left( X, S, V, L \right)
\end{equation} 
where $\omega$ is a function with respect {to} $X$, $S$, $V$ and $L$. Multi-objectives can be represented by multiple functions of $\omega$ as 
\begin{equation}\label{eq2}
    \min \omega=\left\{
    \begin{aligned}
    \omega_1\left( X, S, V, L \right)\\
    \omega_2\left( X, S, V, L \right)\\
    \dots
    \end{aligned}
    \right.
\end{equation} 

For example, the objective of minimizing makespan can be expressed as Eq. (\ref{eq4}) and that of minimizing total running time as Eq. (\ref{eq5}) assuming each node is either idle or processing only one task \citep{z2}.
\begin{equation}\label{eq4}
\min {\omega _{makespan}} = \left( {\mathop {\max }\limits_{j = 1,2, \ldots ,N} \left( {\sum\limits_{i = 1}^M {{x_{ij}}{v^{PT}_{ij}}} } \right)} \right)
\end{equation}
\begin{equation}\label{eq5}
\min {\omega _{total \_ time}} =  \left( {\sum\limits_{j = 1}^N {\sum\limits_{i = 1}^M {{x_{ij}}v^{PT}_{ij}} } } \right)
\end{equation}
where $v^{PT}_{ij}$ means the processing time of the $i$-th task when executed in the $j$-th nodes which belongs to one dimension of $V$ as time can also be regarded as a dimension of resources.
The objective of load balancing can be expressed as Eq. (\ref{eq6}) when using variance of load to measure the degree of balancing \citep{z1}.
\begin{equation}\label{eq6}
\min \omega_{variance} = { \frac{1}{N}\sum\limits_{j = 1}^{N} {{{\left( {\sum\limits_{i = 1}^{M} {{x_{ij}}{v^{DS}_{ij}}} } \right)}^2}}}  
{- { {\frac{1}{N^2}}\left(\sum\limits_{j = 1}^{N} {\left( {\sum\limits_{i = 1}^{M} {{x_{ij}}{v^{DS}_{ij}}} } \right)}\right)^2 }}
\end{equation}
where $v^{DS}_{ij}$ means of disk storage requirement when the $i$-th task is allocated to the $j$-th node  which also belongs to one dimension of $V$.

Assuming the power of the $j$-th node at time $t$ is related to the load status $L_j(t)$, marked as $P_j(t)=P_j \left(L_j(t) \right)$, thus the objective of minimizing energy consumption of the whole system from time 0 to $T$ can be written as Eq. (\ref{eq7}) \citep{b58,b190,ln6}.
\begin{equation}\label{eq7}
\min \omega_{energy\_consumption} = \sum\limits_{j = 1}^N {\int_0^T {{P_j}\left( {{L_j}\left( t \right)} \right)} } 
\end{equation}

From the above examples, most of the optimization objectives in Cloud scheduling can be expressed by the structure of $\omega\left( X, S, V, L \right)$.

\begin{figure}[ht]
    \centering
        
            \includegraphics[width=0.8\columnwidth]{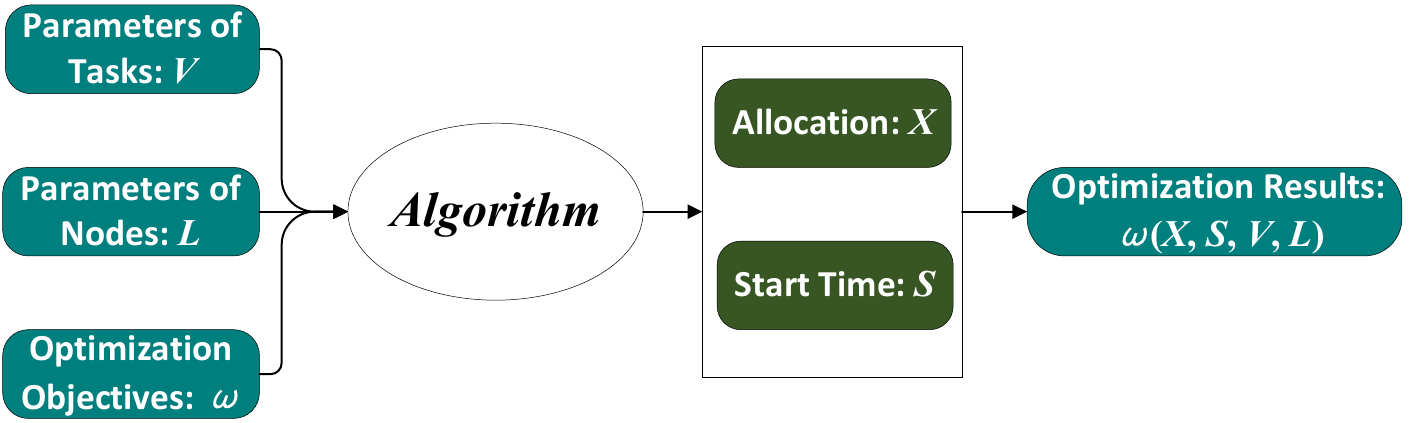}
            \centering
            \caption{A Diagram of Scheduling Algorithm to Generate the Scheme $\left\langle X, S \right\rangle$.}
            \label{figadd1}
      
         \end{figure}

With the formulation of the scheduling problem, a scheduling algorithm is an integration of mappers from $\left(V,L,\omega\right)$ to the scheduling scheme $\left\langle X, S \right\rangle$. It can be set an algorithm as $Al$ and its solution can be expressed by
\begin{equation}\label{eq3}
    Al\left(V,L,\omega\right)=\left\langle X, S \right\rangle
\end{equation}
Thus, a process of using an algorithm to solve the optimization solutions can be shown as Fig.  \ref{figadd1}.
From Fig.  \ref{figadd1}, two key factors for scheduling are production and evaluation of schemes. In solving scheduling schemes, the evaluation for the performance of an optimized solution, i.e. the process of obtaining $\omega\left( X, S, V, L \right)$ or its equivalent evaluation functions, is crucial. Some simple optimization objectives in ideal scenarios can be directly calculated.
However, for some complex optimization objectives,  this function $\omega\left( X, S, V, L \right)$ may not have explicit expressions. {For example}, for minimizing energy consumption in Eq. (\ref{eq7}), $P_j \left(L_j(t) \right)$ cannot be represented by elementary functions generally so that the expression of $\omega\left( X, S, V, L \right)$ is implicit.
For some optimization objectives with explicit expressions in ideal scenarios, it may be also difficult to directly calculate the optimization results when in some highly stochastic system processes. {For example}, when the processing time $v^{PT}_{ij}$ in Eq. (\ref{eq4}) is random {and} satisfying different distributions, the makespan will also be random. Thus, the selection of scheduling algorithms should be based on the characteristics of scenarios. 
The different mapping processes of Eq. (\ref{eq3}) will correspond to different categories of algorithms.

\begin{figure}[ht]
    
            \centering
            \includegraphics[width=0.9\columnwidth]{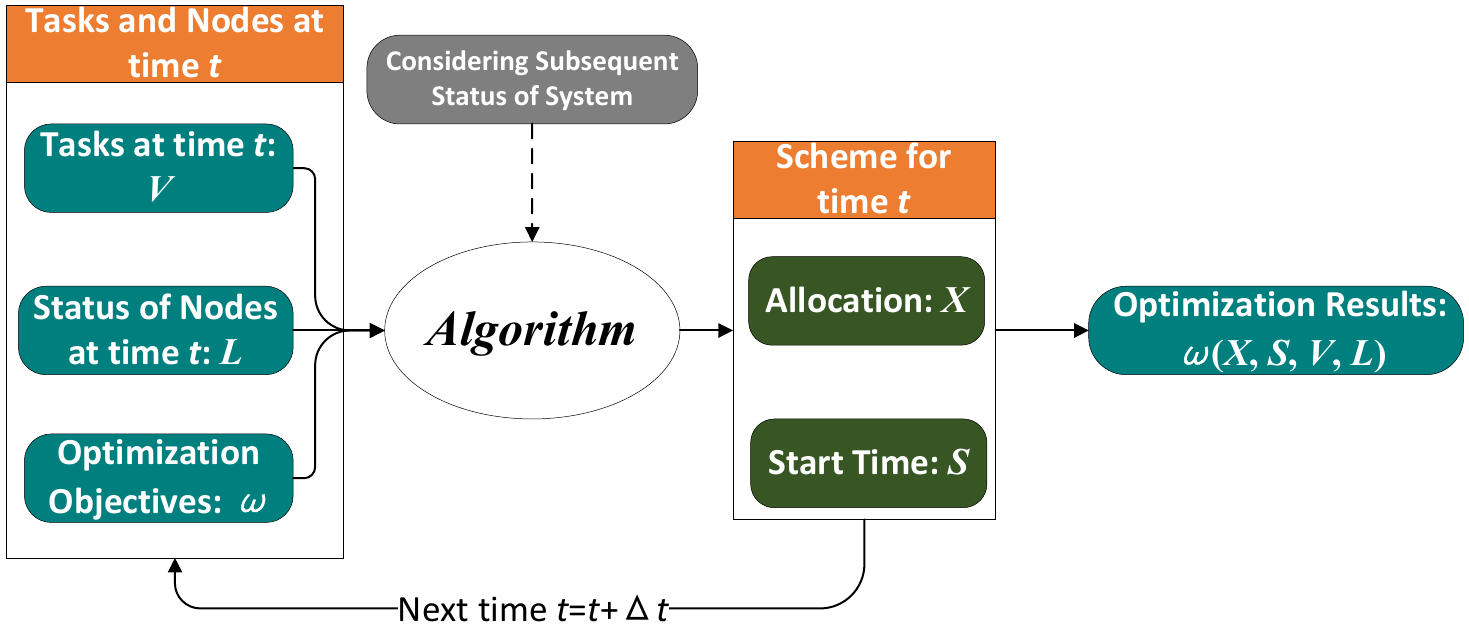}
            \centering
            \caption{A Diagram  for Continuous Dynamic Scheduling Process over Time $t$.}
            \label{figadd3}
         \end{figure}
         
When considering dynamic scheduling, a diagram of its process over time can be seen in Fig.  \ref{figadd3}. 
The scheduling scheme at a time $t$ is responsible for meeting the scheduling requirements at the current time, but will also be related to the status of server nodes at subsequent times. It indicates that when making scheduling decisions at time $t$, it is necessary to consider the subsequent changes in the system. This also puts forward requirements for evaluating the quality of scheduling schemes, which shows the significance of a predictor.

Generally, algorithms for Cloud scheduling contain six categories: Dynamic Programming (DP), Probability algorithm (Randomization), Heuristic method, Meta-Heuristic algorithm, Hybrid algorithms and Machine Learning. From the properties of these algorithms, except for ML, other algorithms do not have the ability to predict system states. In this paper, we regard dynamic programming, randomization, heuristic method, meta-heuristic algorithm, and hybrid algorithm as classic approaches. In order to analyze the future direction of Cloud scheduling and discuss the potential application of DRL, we will review the current scheduling algorithms of Cloud.

\subsection{Review for Classic Algorithms}\label{sec4.1}

For classic approaches, the most commonly utilized methods in surveyed literature are heuristic, meta-heuristic and hybrid algorithms. Thus, we mainly review these three types of algorithms to assist the later review and discussion on the application of DRL in Cloud scheduling.

\subsubsection{Heuristic Algorithms}
Heuristic is an algorithm to solve an optimization problem based on intuitionistic or empirical construction. 
Due to their lower complexity, heuristic algorithms are prevalent in some scenarios with a clear evaluation function requiring rapidity but not requiring high optimization results. Additionally, the worst-case of heuristic algorithms is generally predictable hence with a lower risk of improper allocation.

In existing research, \cite{b87}  applied the Jacobi Best-response Algorithm (JBA) to minimize cost in Multi-Broker Mobile Cloud Computing Networks and proved theoretical results demonstrating the existence of disagreement points and convergence of Jacobi Best-response Algorithm of the brokers to disagreement points.
\cite{b120} proposed an adapting Johnson's model-based algorithm with 2-competitive to minimize the makespan of multiple MapReduce jobs and proved its performance in theory.
{\cite{ln2} proposed Peak Efficiency Aware Scheduling (PEAS) to optimize the energy consumption and QoS in the on-line virtual machine allocation and reallocation of Cloud.}
Dynamic Bipartition-First-Fit (BFF), a $(1+\frac{g-2}{k}-\frac{g-1}{k^2})$ competitive algorithm based on First-Fit algorithm, was proposed and its performance was proved theoretically by \cite{b62}.
\cite{b141} proposed a QoS-Aware Distributed Algorithm based on first-come-first-improve (FCFI) and all-come-then-improve (ACTI) algorithms to reduce computation time and energy consumption of Industrial IoT-Edge-Cloud Computing Environments.
{ECOTS (energy consumption optimization cloud task scheduling algorithm), with low time and space complexity, took into account multiple key factors such as task resource requirements, server power efficiency model and performance degradation in order to reduce
energy consumption of Cloud \citep{ln3}.}
Longest Loaded Interval First algorithm (LLIF), a 2-approximation algorithm with theoretical proof of its performance, was proposed by \cite{b68} to minimize the energy consumption of VM reservations in the Cloud.

Other common heuristic methods are Johnson's model, FF (first fit), BF (best fit), RR (round-robin), FFD (first fit decreasing), BFD (best fit decreasing), Jacobi Best-response Algorithm \citep{b87} and their variants.

\begin{table*}
\caption{A Summary of Heuristic Algorithms.}
\label{table4}

\setlength{\tabcolsep}{2pt}
\begin{tabular}[ht!]{p{95pt}p{80pt}p{90pt}p{80pt}}
%\begin{longtable}%最多380
\hline
Heuristic Algorithm & Scenario &Sever Nature & Objectives\\
\hline

JBA  \cite{b87}& Dynamic scheduling & Heterogeneous servers & Cost \\
EWBS  \cite{b127}& Dynamic scheduling & Heterogeneous servers & Reliability\\
FISTA  \cite{b116}& Dynamic scheduling& Heterogeneous servers &Load Balancing\\
HScheduler  \cite{b120}&Dynamic scheduling& Homogeneous servers&Makespan \\
LARAC  \cite{b63}& Mobile Cloud& Delay-tolerant tasks&Energy consumption\\
{PEAS \cite{ln2}}& Online scheduling& Heterogeneous servers & Energy, QoS \\
Bipartition-First-Fit  \cite{b62} &Online scheduling& Homogeneous servers & Energy consumption\\
MSNWF \cite{b136}&Dynamic scheduling&Heterogeneous C-RAN&Energy consumption\\
FCFI+ACTI  \cite{b141}&Task offloading& Heterogeneous servers&Energy consumption\\
{ECOTS \cite{ln3}} & Static scheduling& Heterogeneous servers& Energy consumption\\
LLIF \cite{b68} &Static scheduling&Heterogeneous server&Energy consumption\\
LPT \cite{b501}&Static scheduling&Homogeneous servers&Makespan \\
\hline
\end{tabular}
\end{table*}

To give an overall observation, we collected the reviewed literature and gained Table \ref{table4}. From Table \ref{table4},  
heuristic algorithms mainly focus on the {single-objective} optimization including minimizing makespan, minimizing energy consumption and load balancing. However, there are several defects of heuristics as follows.
\begin{enumerate}[label=(\arabic*), leftmargin=*, topsep = 0em, partopsep=0em, itemsep=0em, parsep = 0em]
  \item For the scenarios using heuristic, some major objects (such as the time, energy or load) are often assumed to be given or easily calculated. For complex scenarios where the optimization objective is implicit with respect {to} solutions, heuristic algorithms often fail to generate a feasible solution.
  \item A heuristic algorithm is often designed for one or few specific scenarios. When only one element in the scenario changes, the algorithm may need to be redesigned.
  \item Heuristics are usually only suitable for the {single-objective problems}.
  \item Moreover, the solution of heuristic algorithms can usually be further optimized.
\end{enumerate}

\subsubsection{Meta-Heuristic Algorithms}
In skeleton, meta-heuristic, the combination of heuristic and randomization \citep{b76}, includes Ant Colony Optimization (ACO), Particle Swarm Optimization (PSO), Artificial Bee Colony (ABC), Genetic Algorithm (GA), Firefly Algorithm (FA), etc.

ACO imitates ant colony to search for food as a search
route.
\cite{b126} proposed OEMACS combining OEM (order exchange and migration) local search techniques and ACO to resolve energy consumption of VMs deployment in Cloud computing, which significantly reduced the energy consumption and improved the effectiveness of different resources compared with conventional heuristic and other evolutionary-based approaches.
\cite{b137} proposed two ant colony-based optimization algorithms (TACO) to address VM scheduling and routing in multi-tenant Cloud data centers aiming at improving the utilization of energy in Cloud computing.
\cite{b85} proposed an alternative meta-heuristic technique based on the Ant Lion Optimizer Algorithm (MALO) to resolve multi-objective optimization of Cloud computing, which performed better in load balancing and makespan compared with GA, MSDE, PSO, WOA, MSA and ALO.

GA imitates the process of natural evolution as a search route of the algorithm.
Proposed by \cite{ln7}, NSGA-II occupies better convergence and optimal solution and has become one of the benchmarks using the fast non-dominated sorting algorithm, introducing elite strategy and using congestion-congestion comparison operator. \cite{b130} improve the search strategy based on NSGA-II to reduce the energy consumption, response time, load imbalance and makespan in Cloud computing.
NSGA-III utilizes reference points with preferable distribution as a novel search route to maintain the diversity of the population to improve the optimization results of GA \citep{ln8,b303}.
\cite{b142} applied NSGA-III to optimize the execution time and energy consumption of IoT-enabled Cloud-edge computing. MOGA \citep{b131} and MOEAs \citep{b56} improved the search route strategies based on NSGA-II and were utilized to settle Cloud scheduling.

The studies of Firefly algorithm include FA \citep{b84} and FIMPSOA \citep{b117}. That of PSO include MOPSO \citep{b65}, TSPSO \citep{b134} and HAPSO \citep{b125}. Other meta-heuristic algorithms include  Multi-objective Whale Optimization Algorithm (MWOA) \citep{b147}, nature-inspired Chaotic Squirrel Search Algorithm (CSSA) \citep{b157}, etc.

%Improved differential evolution algorithm(IDEA) \cite{b133}, Multi-heuristic resource allocation algorithm (MHRA) \cite{b64}, Efficient priority and relative distance (EPRD) algorithm \cite{b73}, Multi-objective evolutionary list scheduling (MOELS) algorithm& \cite{b145}, }.

\begin{table*}[ht!]
\caption{A Summary of Meta-Heuristic Algorithms.}
\label{table3}
\setlength{\tabcolsep}{2pt}
\begin{tabular}{p{60pt}p{105pt}p{190pt}}

%\begin{longtable}%最多380
\hline
Subcategories & Algorithm &Objectives \\
\hline

\multirow{7}{*}{ACO}
%&ACO &  \cite{b126}\\%b77,
 &MALO \cite{b85}& Makespan\\
 & HGA-ACO \cite{b106}& Makespan\\
 & DAAGA  \cite{b160}& Running time, QoS\\
 & TACO \cite{b137}& Energy consumption \\
 & OEMACS  \cite{b126}&Utilization\\
 & S-MOAL  \cite{b135}& Energy consumption  \\

\hline
\multirow{7}{*}{GA}
&NSGA-II \cite{b74}& Makespan, Energy consumption\\
&TS-NSGA-II  \cite{b130}& Running time, Utilization \\
%&Parallel multi-objective genetic algorithm \cite{b162}& Cost, Load balance&\\
&MOGA \cite{b57,b124}&Makespan, Cost\\
&NSGA-III   \cite{b142}& Energy consumption\\
&MOEA/D-based GA \cite{b123,b56}& Qos: Cost, Queue time \\

\hline
\multirow{4}{*}{PSO}
&MOPSO  \cite{b65,b96}&  SLA violations, Energy consumption \\
&TSPSO  \cite{b134}&Running time, Energy consumption, Failed Task\\
&HAPSO  \cite{b125}&Response time, Energy consumption\\
&PSO-based MOS  \cite{b78}&Cost, Makespan\\
\hline

\multirow{2}{*}{FA}
&FA \cite{b84}&Makespan, Utilization\\
&FIMPSOA \cite{b117}& Utilization, Reliability, Throughput\\
\hline

\multirow{2}{*}{Others}
&MWOA \cite{b147}&Cost, Utilization, Energy consumption\\
&CSSA \cite{b157}&Cost, Time, Energy consumption, Utilization\\
\hline

\end{tabular}
\end{table*}

By collecting and sorting out the literature using meta-heuristic algorithms to solve resource scheduling problems, we gain Table \ref{table3} with their corresponding optimization objectives.
Since the meta-heuristic algorithms are also applicable to the scenarios of dynamic scheduling and heterogeneous servers where the heuristic algorithms are applicable, their application scenarios are not listed in Table \ref{table3}.
From Table \ref{table3}, meta-heuristic algorithms with searchability for the solution can address more complex optimization problems not only for single-objective problems but also for multi-objective problems. 
They are applied to solve optimizing cost, energy consumption, makespan, running time and resource utilization. Meta-heuristic algorithms are more applicable than heuristic algorithms but at the expense of computational complexity and randomness. However, although these optimization objectives in meta-heuristics include some complex objects (such as energy, Qos {and} cost), their calculations have been simplified with some ideal assumptions far from reality  \citep{b137,b124,b142}.

\begin{figure}[ht]
    \centering
        
            \centering
            \includegraphics[width=\columnwidth]{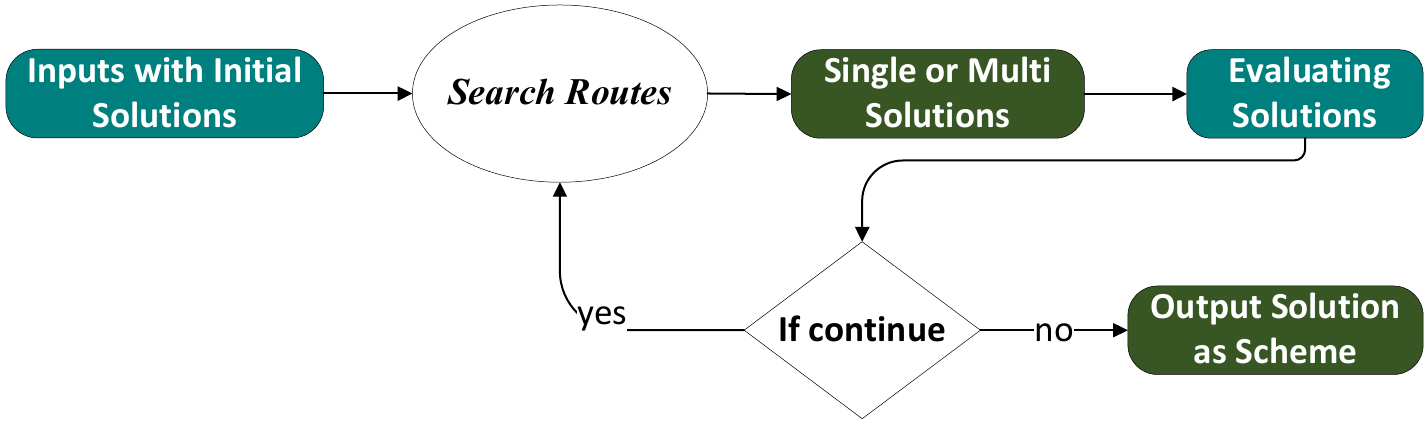}
            \centering
            \caption{A Diagram of Search-based Algorithms.}
            \label{figadd2}
      
         \end{figure}
Meta-heuristic and other search algorithms are based on the specific search route, whose diagram can be seen in Fig. \ref{figadd2}.
They use the search route to adjust the current solutions to generate new solutions, evaluate the performance (such as fitness) of newly generated solutions according to the optimization objectives-based evaluation functions, and then determine whether to proceed to the next search based on the evaluation. The two key factors in Fig. \ref{figadd2} are the search route and evaluation of solutions. The search route needs to generate better solutions. However, there are several inevitable defects of meta-heuristic as follows.
\begin{enumerate}[label=(\arabic*), leftmargin=*, topsep = 0em, partopsep=0em, itemsep=0em, parsep = 0em]
  \item For scenarios where the solution can be directly evaluated, the convergence of the meta-heuristic cannot be guaranteed due to the presence of randomness. The randomness of the meta-heuristic increases redundant computations. 
  \item As the search space increases, the required search iterations must also increase accordingly, subsequently producing more redundant solutions.
  \item When it is difficult to evaluate the quality of a solution, the search route will lose its direction, and the search algorithm will degenerate into pure randomization.  When $\omega\left(X,S,V,L \right)$ is implicit, the meta-heuristic and other search algorithms themselves do not provide a method for evaluating solutions.
  \item Meta-heuristic also does not provide a way to predict system status.
\end{enumerate}
The first and second defects will limit the optimality of the meta-heuristic for its feasible scenarios.  The third and fourth defects, which also appear in heuristic, will cause the algorithm unable to be used in some real-world complex scenarios.

\subsubsection{Hybrid Algorithms}
Some other classic algorithms used in Cloud scheduling mainly contain DP, Random algorithms, and hybrid algorithms (combining two or more algorithms).
Among them, hybrid algorithms are also widely used in solving complex scheduling problems in Cloud computing.
Hybrid algorithms can combine the advantages of multiple algorithms to produce better solutions. In terms of search algorithms, a single algorithm has an inherent local convergence solution and the solution of the hybrid algorithm needs to satisfy the convergence conditions of multiple algorithms simultaneously \citep{z2}. Therefore, the convergence solution of the hybrid algorithm is usually better than the corresponding single algorithm.
PSO-ACS \citep{b251}, mingled with PSO and ACO, applied PSO to find the optimal solution of task scheduling and ACO to find the best migration path of VMs on PMs.
FACO \citep{b258}, a hybrid fuzzy ant colony optimization algorithm, exploited a fuzzy module dedicated to pheromone evaluation to improve the performance of ACO by optimizing the search route of ACO.
Hybrid Genetic-Gravitational Search Algorithm (HG-GSA) \citep{b264} based on gravitational search algorithm for searching the best position of the particle consequently optimizing the search route of GA.
FMPSO (modified PSO + fuzzy theory) \citep{b265} used crossover and mutation operators surmounting the local optimum of PSO and applied a fuzzy inference system for fitness calculations.
SFLA-GA algorithm (shuffled frog leaping algorithm + GA) \citep{b266} took advantage of the two algorithms to transmit information among groups hence the optimal search route.
GHW-NSGA II \citep{z3} leveraged heuristic-based search algorithm as an extra search route of NSGA II to optimize the utilization of multi-dimensional resources, which improved the convergence speed and optimality on the basis of GA.
SPSO-GA \citep{n10} combined Self-adaptive Particle Swarm Optimization algorithm with Genetic Algorithm operators to reduce the energy consumption of the scenario offloading DNN layers Cloud-Edge environment. On the basis of SPSO-GA, PSO-GA-G \citep{n11} added a greedy algorithm to optimize computation offloading. The combination of multiple meta-heuristics is beneficial for improving the overall convergence speed, hence improving search efficiency.
LPT-One and BFD-One \citep{z2} used heuristic algorithms to act as the search routes and combined multiple heuristic-based search routes to improve the approximation of minimizing makespan.

Other hybrid algorithms in Cloud scheduling, include ABC-SA integrating the functionality of simulated annealing (SA) into artificial bee colony \citep{b267}, SFGA (a hybrid Shuffled Frog Leaping Algorithm and Genetic Algorithm) \citep{b263}, TSDQ-hybrid meta-heuristic algorithms based on Dynamic dispatch Queues \citep{b270}, etc. These algorithms demonstrated the flexibility, superiority, adaptability and mobility of hybrid algorithms and simultaneously manifested the unlimited possibilities and significance of research hybrid structurally.

Similar to meta-heuristic algorithms, hybrid algorithms are also applicable to a variety of multi-objective problems. However, a hybrid algorithm, with multiple heuristics or meta-heuristics as elemental algorithms, cannot exceed the scenarios that the elemental algorithms are suitable for, which implies that it is also not suitable for the scenarios with implicit $\omega\left(X,S,V,L \right)$.

\subsubsection{Summary of Classic Algorithms}
Although the classic algorithms have been applied to various objectives under various scenarios and achieved considerable performances, they still do not solve how to calculate or evaluate the various elements such as energy, time, load and utilization according to the properties of tasks and resources.
Therefore, the models of Cloud computing in their applications are different from the realistic scenes, which causes them to only be applicable when the elements (such as time, cost, energy, and load) are given or easy to calculate. This also leads to the difference between the expected performance of these algorithms and the actual operational performance in reality. When the mapping of objective $\omega\left(X,S,V,L \right)$ is implicit, classic algorithms are unable to guarantee optimality and are even unable to obtain feasible solutions. This is because classic algorithms do not provide a method to measure the performance of solutions. 
In addition, for a new optimization problem, classic algorithms, without memorability, need to resolve the optimization solution from scratch.

\subsection{Machine Learning}\label{sec4.2}

Before providing {a} detailed introduction {to} DRL-based algorithms in Cloud scheduling, we give a collection of ML methods used in Cloud scheduling  by reviewing literature in Table \ref{table6}. The ML methods used in scheduling problems mainly contain deep learning (DL), RL and DRL. {In addition,  other types of machine learning methods, such as KNN \citep{z4, ln1} and imitation learning \citep{b107, b95}, SVM \citep{ln6, b92}, had also been applied in cloud scheduling.}

\begin{table}[ht!]

\caption{Machine Learning Methods in Cloud Scheduling.}
\label{table6}
\setlength{\tabcolsep}{1pt}
\begin{tabular}{p{60pt}p{110pt}p{50pt}}

\hline
Subcategories & Algorithms & {References}  \\
\hline

\multirow{2}{*}{DL }
&NN-DNSGA-II algorithm &\cite{b82}\\
&DLSC framework &\cite{b111}\\
\hline

%\hline
\multirow{11}{*}{\shortstack{RL }}
&QEEC &\cite{b58}\\
&RLVMrB &\cite{b112}\\
&RL+Belief-Learning &\cite{b167}\\
&Revised RL &\cite{b170}\\
&URL &\cite{b171}\\
&Q Learning Algorithm &\cite{b172}\\
&Bare-Bones RL &\cite{b173}\\
&Adaptive RL &\cite{b174}\\
&Rethinking RL &\cite{b175}\\
&ML+RM  &\cite{b201}\\
&RL-based ADEC &\cite{b203}\\
\hline

\multirow{18}{*}{\shortstack{DRL}}
&DERP  &\cite{b149}\\
&A3C RL algorithm &\cite{b150, b113,n12}\\
&RL-based DPM framework &\cite{b177}\\
&Deep Q Network (DQN) &\cite{b152, b164,n3}\\
&ADRL &\cite{b169}\\
&DeepRM-Plus &\cite{b107}\\
&PCRA &\cite{b13}\\
&Modified DRL algorithm &\cite{b153}\\
&DQTS &\cite{b83}\\
&MRLCO &\cite{b161}\\
&Multiagent DRL &\cite{b168}\\
&DL$^2$ &\cite{bd101}\\
&DRL+FL &\cite{b190}\\
&RLFTWS &\cite{n1}\\
&AV-MPO &\cite{n2}\\
&HCDRL &\cite{n4}\\
& DT  &\cite{n5}\\
& CORA &\cite{n6}\\
&  DRAW &\cite{n7}\\

%\textcolor[rgb]{1.00,1.00,0.00}{Other}&\textcolor[rgb]{1.00,1.00,0.00}{Multi-Agent Imitation Learning \cite{b95}}\\

\hline
\end{tabular}
\end{table}

In practice, a Cloud system has several characteristics:
\begin{itemize}[leftmargin=*, topsep = 0em, partopsep=0em, itemsep=0em, parsep = 0em]
\item large scale and complexity of systems that make it impossible to model accurately;
\item timeliness of scheduling decisions that demands the high-speed scheduling algorithm;
\item randomness of tasks (or requests) including randomness of task numbers, arrival time and sizes.
\end{itemize}
These characteristics are challenging for the research on Cloud scheduling. Most existing
optimization methods are designed for specific applications \citep{ln1}. When we constantly consider more factors in the process of modeling Cloud scheduling, the existing classic algorithms are no longer applicable. It is tough for one specific meta-heuristic, heuristic and hybrid algorithms to fully adapt to the real dynamic Cloud computing systems or Edge-Cloud computing systems.

Considering the defects of classic algorithms, ML-related methods can utilize specific mapping methods to record the computational mode of optimization objectives. This addresses the dilemma of evaluating the quality of a solution when $\omega\left(X,S,V,L \right)$ is implicit. E.g., the Q-table in Q-learning and various DNNs in DRL both embed the ability to evaluate the quality of a solution with some memorability.
While, there is no given target scheduling scheme as the label, which makes it impossible to solve the scheduling problem solely using DL. One effective approach is to apply DL in meta-heuristic to evaluate the quality of solutions using the realistic situation of the system to obtain optimization objectives and to train neural networks. This enables meta-heuristics to perform effective searches, such as NN-DNSGA-II algorithm (combining DL with GA)  \citep{b82} and DLSC framework (combining DL with PSO)  \citep{b111}.

A novel type of ML policy for Cloud scheduling is the combination of DNN and RL, called deep reinforcement learning. Different from the combination of DL and meta-heuristic, DRL leverages DNN to act as the solution generator.
Integrating the advantages of RL and DL, DRL has the following benefits.
\begin{itemize}[leftmargin=*, topsep = 0em, partopsep=0em, itemsep=0em, parsep = 0em]
\item Ability of modeling: it can model complex systems and decision-making policies with DNN even when $\omega\left(X,S,V,L \right)$ is implicit;
\item Adaptability for optimization objectives: training progress based on gradient descent algorithm makes it possible to search the optimization solution for various objectives;
\item Adaptability for the environment: DRL can adjust parameters to adapt to various environments;
\item Possibilities for further growth: DRL can grow over time to process large-scale tasks;
\item Adaptability for state-space: DRL can process continuous states or multi-dimensional states;
\item Memory of experience: DRL possesses the capacity to memory experience with experience replay.
\end{itemize}

For the sake of demonstrating the above benefits and further analyzing the challenges of DRL in scheduling, the next section will introduce and analyze the framework of RL and DRL as the foundation to support the follow-up review and discussion.

\section{Analysis of RL and DRL Frameworks in Scheduling}\label{sec5}

Machine learning is the discipline of teaching the computer to predict outcomes or classify objects without explicit programming \citep{b92}.
ML is also an artificial intelligence discipline of studying how computers simulate or implement human learning behavior so that computers can gain new knowledge and skills.
Based on learning methods, ML can be divided into supervised learning, unsupervised learning and semi-supervised learning \citep{b92}.
RL is one of the unsupervised learning \citep{b203}. Based on learning strategy, ML contains Symbol learning, Artificial Neural Networks learning, Statistical ML, Bionic ML, etc.
DL on the basis of deep artificial neural networks and RL are two subsets with the intersection of ML, where the intersection between DL and RL is DRL. DRL combined with the perception of DL and with the decision-making ability of RL, has been applied in robot control, computer vision, natural language processing and some Go sports \citep{b196,b200}.

From the scheduling formulation and classic algorithms in {Sect.} \ref{sec4},  the two key factors of scheduling are the production and evaluation of solutions. The classic algorithms, including heuristics and meta-heuristics, don't {possess} the ability to evaluate solutions or {to} predict the status of the system when $\omega\left(X,S,V,L \right)$ is implicit. Therefore, they suffered in some realistic scenarios.
Due to the combination of DL and RL, DRL has the flexibility of adopting DNN to complete any process in solving the scheduling schemes. Meanwhile, the RL mechanism in DRL maintains the well-performed solution, as it ensures statistical advantages of performance through training. RL in DRL, based on the theory of the Markov process, is also suitable for dynamic scheduling \citep{n1,n2,n4}.

The application of DRL to resolve the scheduling problems in Cloud computing emerged in recent years, which is an effective intersection of two emerging technologies. DRL has shown superior performance in the current research on the application in Cloud scheduling. 
To support subsequent review of existing research and discussion of challenges and future directions, we introduce and analyze the evolution of RL and DRL frameworks in this section, which can provide a comprehensive insight into understanding the operation process of DRL.

\subsection{RL Framework}\label{sec5.1}
\begin{figure}[ht]
    \centering
        \begin{minipage}[h]{0.70\columnwidth}
            \centering
            \includegraphics[width=\columnwidth]{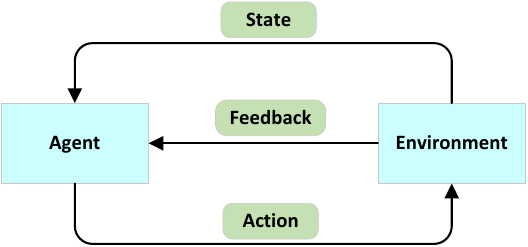}
            \centering
            \caption{A Fundamental Framework of RL.}
            \label{fig1}
        \end{minipage}
         \end{figure}

Firstly, we introduce and analyze the framework of RL.
RL is based on the interaction model between the agent and the environment. It instructs the agent to learn the optimal action strategy by the feedback from the environment corresponding to the agent's action. The RL model can update the action strategy according to timely feedback and long-term feedback. The agent will choose the action on the basis of the action strategy.
State-space, action-space, environment, feedback (reward), and strategy are five basic elements of RL. Figure \ref{fig1} shows a fundamental structure of RL with these basic elements.

In some of the reviewed literature related to RL \citep{b164,b169,b177}, the feedback is described as a reward.  In this paper, we regard it as feedback considering that both positive feedback and negative feedback will affect the learning progress of the agent's action strategy \citep{b83,b107,b151}.
The concept ``feedback'' originated from cybernetics \citep{n14}. Based on the perspective of feedback in RL, RL requires feedback from the environment outside of the solver, while meta-heuristics and other search algorithms typically rely on evaluation functions; the role of feedback in RL is to update the solver (i.e., affecting how to solve a scheme) \citep{n15,n16,b17,n18,b97}, while evaluation functions in meta-heuristics or other search algorithms are responsible for update the solution (directly changing the scheme). The fundamental framework of RL in Fig. \ref{fig1} provides a simplified overall structure, but it is not sufficient to directly solve scheduling problems. Especially, this architecture does not solve the difficulties that we need to face when solving the scheduling problems mentioned above. Therefore, we need to further evolve the architecture of RL based on this foundation framework.

\begin{figure}[ht!]
         \centering
        \begin{minipage}[h]{0.99\columnwidth}
            \centering
            \includegraphics[width=\columnwidth]{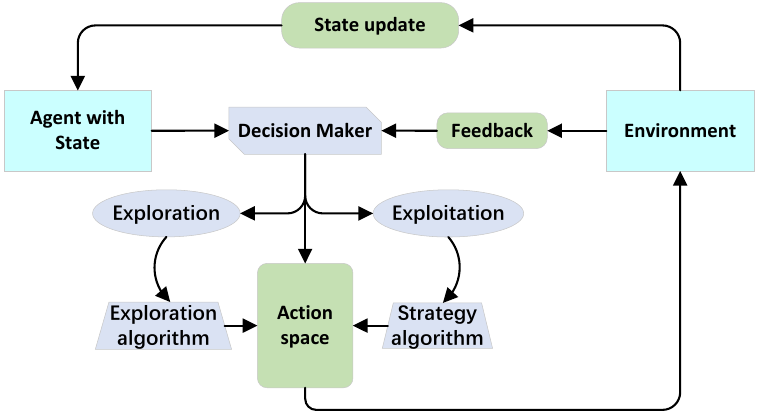}
            \centering
            \caption{A Framework of RL with Action Selection and Strategy Update.}
            \label{fig2}
        \end{minipage}
        \end{figure}

In another standpoint to comprehend the fundamental structure of RL, the agent learns the strategy by trailing error. And trailing error requires the agent to maintain the balance between exploration and exploitation.
The greedy method, random method and meta-heuristic method will be used to simulate the decision progress between exploration and exploitation. Markov decision process (MDP) is a common model to express the action choice process and Bellman Equation, a dynamic programming equation, is a common function to update the action strategy. Hence, a framework of RL containing action selection and strategy update can be shown as Fig. \ref{fig2}.

The framework of RL with action selection and strategy update in Fig. \ref{fig2} can already be used to solve some optimization problems, but has no enough consideration to the temporal changes in the system state and agent state. Thus, it is not sufficient to solve the time-related scheduling problem.

\begin{figure}[ht!]
\centering
        \begin{minipage}[h]{0.99\columnwidth}
            \centering
            \includegraphics[width=\columnwidth]{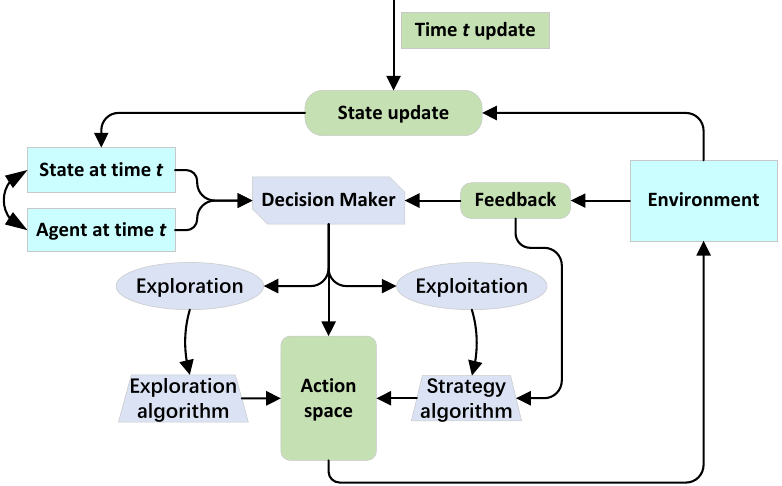}
            \caption{A Complex Framework of RL based on Varying Agent-state.}
            \label{fig3}
        \end{minipage}
          \end{figure}

In complex scenarios, state and agent vary with time. In addition, the decision should be made according to the state and agent in real-time and the feedback from the environment will affect strategy directly. Hence, an agent-state-based structure of RL can be shown in Fig. \ref{fig3}.

\begin{figure}[ht!]
\centering
        \begin{minipage}[h]{0.99\columnwidth}
            \centering
            \includegraphics[width=\columnwidth]{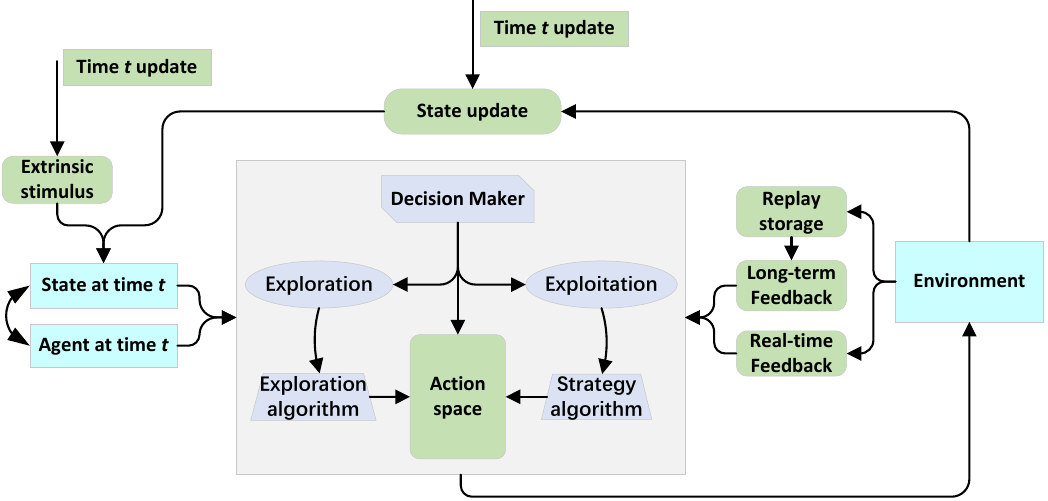}
            \caption{A Framework of RL with Extrinsic Stimulus.}
            \label{fig4}
        \end{minipage}
        \end{figure}

In most realistic scenes, a system is often not completely independent and will alter with extrinsic stimulus.
The environment in Fig. \ref{fig3} is actually the internal environment of the system which cannot express the overall interferences from other systems to this internal system.
Hence, the system of RL in Fig. \ref{fig3} should be regarded as an autonomous system because the agent and environment evolve on the autonomous rules.
A computer game, a Go sport and a language processing problem that covers a large enough amount of data may be regarded as an autonomous system because their regulation is quite stable without external modification of regulation. The movement of vehicles and antagonistic sports are usually affected by external incentives.
Then, a Cloud computing system, with time-varying constructive demands, optimization objectives and users' requests, is not an autonomous system. Moreover, the update function of the internal environment for the agent-state and the decision-making function is also time-varying such as the revenue ratio of Cloud computing is variable in different periods of the same day.
Regarding the decision-making process as an ensemble, a framework of RL with a time-varying extrinsic stimulus can be shown in Fig. \ref{fig4}.

The above frameworks in Fig. \ref{fig1} $\to$ Fig. \ref{fig2} $\to$ Fig. \ref{fig3} $\to$ Fig. \ref{fig4} constitute the evolution process of RL structure from simple to complex. The complicated framework can address a lot of problems in decision-making when the input data is discretized. With increasing complex input data and the increasing dimensions of agent-state, RL frameworks without leveraging DNN are not applicable, because the decision-makers, such as Q-table, are unable to make use of the information of state and feedback without sufficient perception ability, hence the training may not converge. Therefore, it is significant to integrate some neural networks to enhance information perception ability, so as to improve the quality of optimized solutions.

\subsection{ DRL Framework}\label{sec5.2}
\begin{figure}[ht!]
\centering
        \begin{minipage}[h]{0.98\columnwidth}
            \centering
            \includegraphics[width=\columnwidth]{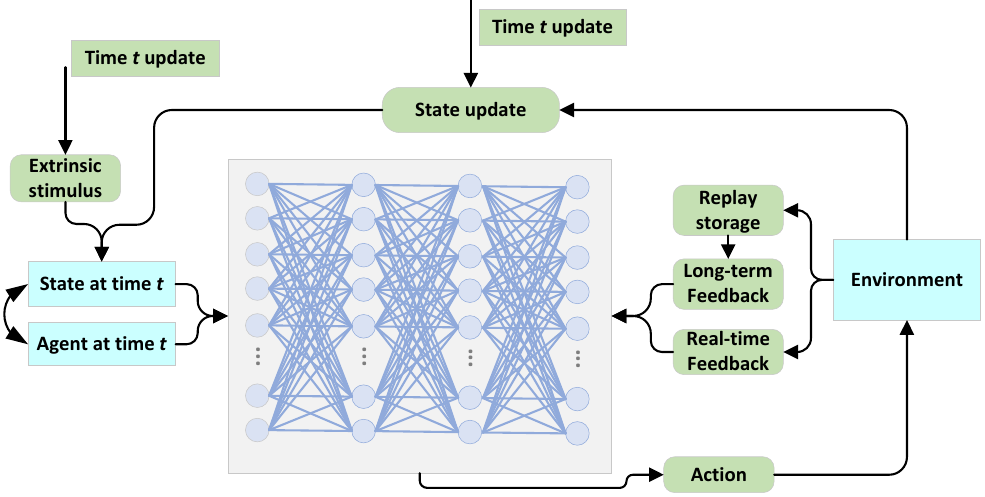}
            \centering
            \caption{A Framework of RL with DNN-based Decision-maker ({Belonging to} DRL).}
            \label{fig5}
        \end{minipage}
        \end{figure}

Before further analyzing the DRL frameworks, we discuss the framework in Fig. \ref{fig4} again on the sight of mapping of mathematics.
In Fig. \ref{fig4}, the decision-maker, which is a complex mapping from agent-state to action, is integrated as an ensemble. Some patterns of RL focus on the expression of this mapping relationship such as Q-table, Advantage Function, Policy Gradient, and Hidden Markov Chain. However, as the sizes and dimensions of state space and action space increase, the computational complexity and storage space of these patterns will grow exponentially.
Furthermore, when the state space is non-discrete which appears in the scheduling problem usually, it is difficult to express the mapping relationship in the general methods of RL.
Nonhomogeneous Markov process-based RL, one of the methods to express the process of time-varying continuous time-space and continuous state space, requires solving differential equations with variable coefficients, however. It astricts the application of the nonhomogeneous Markov process in RL to solve the problem with time-varying continuous time-space and continuous state space.
Hence for various reasons, a DNN with excellent performance in the establishment of mapping relationship is a considerable choice to be a mapper of strategy between state-agent and action. Then, we can improve Fig. \ref{fig4} to a framework using a DNN to express the decision process, shown in Fig. \ref{fig5}.

\begin{figure}[ht!]
\centering
        \begin{minipage}[htbp]{0.90\columnwidth}
            \centering
            \includegraphics[width=\columnwidth]{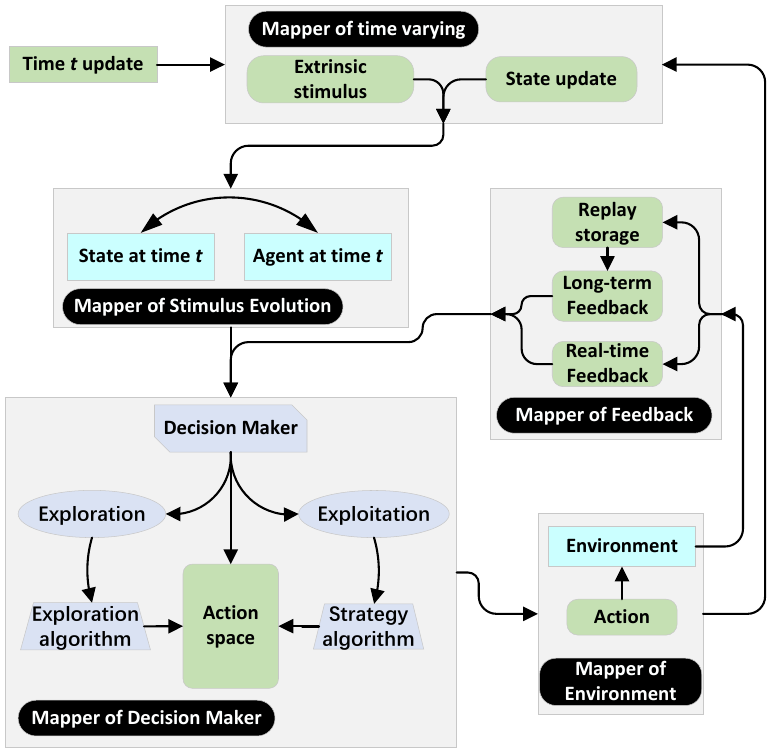}
            \centering
            \caption{A Framework of RL with {Multiple Segments} of System Represented by Multiple Mappers.}
            \label{fig6}
        \end{minipage}
        \end{figure}

The framework in Fig. \ref{fig5} is actually DRL, which can deal with more complex scenarios than  Fig. \ref{fig4} by using DNNs to participate in decision-making.
Regarding the decision process as a mapping process, Fig. \ref{fig5} enlightens us to reconstruct the structure of Fig. \ref{fig4} according to the mapping relation. Then, we can construct the framework in Fig. \ref{fig4} as five mappers including the mapper of time-varying, mapper of stimulus evolution, mapper of decision, environment and mapper of feedback as Fig. \ref{fig6}.

The details of each mapper in Fig. \ref{fig6} are as follows:
\begin{itemize}[leftmargin=*, topsep = 0em, partopsep=0em, itemsep=0em, parsep = 0em]
  \item Mapper of time-varying refers to the relationship between agent-state and time with stimulus from extrinsic or internal space. In it, time and update are the input, as well as stimulus force is the output.
  \item Mapper of stimulus evolution refers to the stimulus evolution in agent and state as agent and state are usually variable with stimulus where stimulus force is the input and the set of agent-state at real-time is output.
  \item Mapper of decision is responsible for calculating the next action according to the current state of the agent where the set of agent-state at this time is input and action at next time is output.
  \item Environment receives actions that the result of the mapper of decision provides, evolves according to the action of the agent and then outputs the environment's state at the next time. The output of the environment enters the mapper of feedback as its input and enters the mapper of time-varying as the internal stimulus for the agent.
  \item Mapper of feedback receives the environment's state, then stores it as replay storage in preparation for long-term feedback in the future and takes it as timely feedback simultaneously. Long-term feedback and real-time feedback will update the parameters of the decision-maker.
\end{itemize}

\begin{figure}[ht!]
\centering
        
            \centering
            \includegraphics[width=0.8\columnwidth]{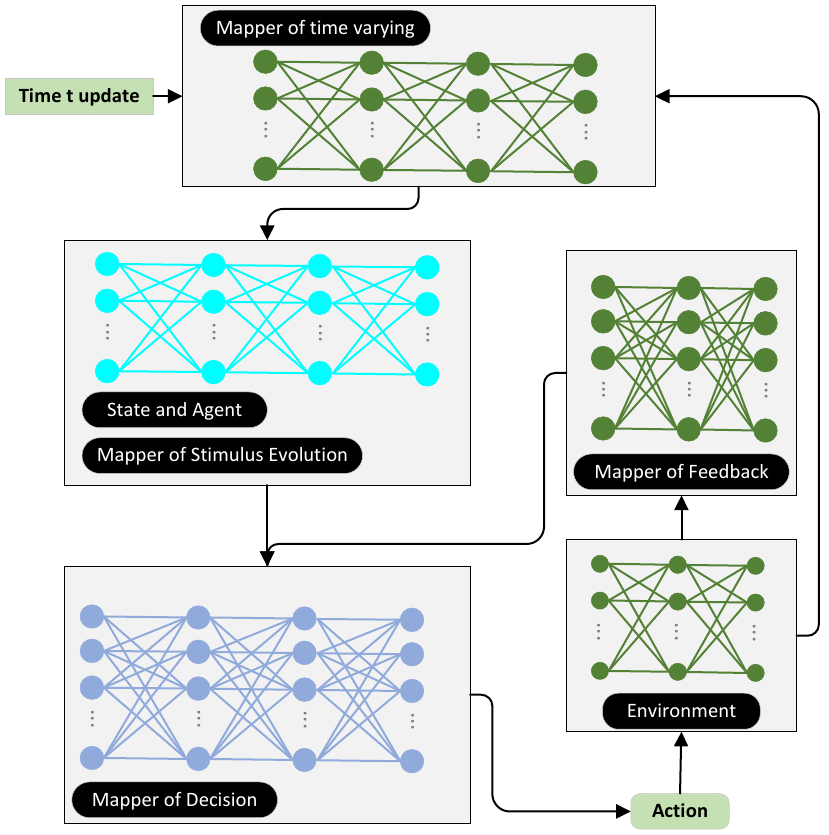}
           
            \caption{ A Framework of DRL with {Multiple} DL Segments of System.}
            \label{figadd4}
    
        \end{figure}

The framework in Fig. \ref{fig6} is a generalized RL based on the integration of mappers. In some of the experiments, the mapper of time-varying, mapper of stimulus evolution and environment are usually simulated by the program or observed in real scenes.
Mapper of decision and mapper of feedback can be constructed with DNNs. As the mapper of feedback is aimed at updating the parameters of the decision-maker, the mapper of feedback can be designed as a neural network to calculate the loss function of the DNN in the mapper of decision, hence Nature DQN or Double DQN \citep{b152,b153,b164,b189} where the neural network of feedback is called as target network with the same structure of decision's network.
While inherently, the five mappers in Fig. \ref{fig6} can all be represented by neural networks respectively. In some scenes, it is difficult to simulate or observe the realistic process of a complex system, and the neural network can be used as an end-to-end alternative.
An extreme example is that five mappers are all expressed with DNNs, shown as Fig. \ref{figadd4}.
However, existing research, using DRL to resolve the scheduling problems in Cloud computing surveyed in this paper, is carried out by replacing one or several of the five mappers with neural networks and has performed well in experiments according to their results, which will be reviewed next section to support the discussion and analysis of the current situation, challenges and future direction of DRL in Cloud scheduling.

\subsection{Summary}\label{sec5.3}
The Cloud environment is a complex and random system with large-scale user requests and a complex physical environment, and these user requests and extrinsic physical environment can be regarded as a time-varying stimulus. In addition, the actual running processes of electronic components and software programs are hard to express using simulation. Crucially, the high dimensionality and the continuity of state-space make the mapper of decision and mapper of feedback difficult to model with conventional methods. In summary, the five mappers may have demand to be modeled with implicit expression functions, while the DNN is a practical method currently to deliver implicit relation based on sufficient data and sufficient training time.

Moreover, the literature adjusts the structure of the neural network (CNN, LSTM, full connection, Transformer, etc.), increases the strategy of initialization of neural network parameters, the training strategy of the neural network, prediction or simulation of the internal and external environment, and assist with other meta-heuristic algorithms as appropriate. With the analysis of the RL framework in this section, we can review and analyze existing DRL-based methods in Cloud scheduling following this perspective.

\section{Overview of DRL-based Scheduling in Cloud}\label{sec6}

Based on the location of the neural network, we review frameworks in surveyed papers. In RL, the central component is the mapper of decisions that can conduct the scheduler of Cloud computing. In Cloud scheduling using RL, the mapper of decision is usually represented by a DNN or Q-table. In order to deeply analyze and macroscopically summarize the application of DRL in Cloud scheduling, we organize the information of literature structurally. In addition to the information of the literature, we also reorganize the possible future work of some literature to provide another probability through the analysis of the reviewed literature.

\subsection{Literature Review}

QEEC \citep{b58} is a Q-learning-based task scheduling framework for energy-efficient Cloud computing using a Q-value table to express the decision-maker of action.

PCRA \citep{n13} is a Prediction-enabled feedback Control with RL-based resource Allocation using a feedback control Q-value prediction model to predict the values of management
operations at different system states.

The DeepRM\_Plus \citep{b107} uses a neural network that has a convolution neural network (CNN) of six layers to describe the mapper of the decision based on the great success of DNN in image processing. The center cluster, waiting for queues, and backlog queue compose the state of the environment which is represented by an image.

AGH+QL \citep{b170}, a novel revised Q-learning-based model, takes hash codes as input states with reduced size of state space.

DQST \citep{b83}, deep Q-learning task scheduling, uses the fully connected network to calculate the Q-values which can express the mapper of action decision.

DERP \citep{b149} uses three different approaches of a DRL agent to handle the multi-dimensional state and to provide elastic VM resources.

Modified DRL \citep{b153}, RLTS \citep{b164}, DRL-Cloud \citep{b189}, and ADRL \citep{b169} also use the structure of action-value Q network (or called evaluate Q-network \citep{b164}) and target-Q network. Then, their similarities and differences are as follows.

IDRQN \citep{b151} is a fine-grained task offloading scheme based on DRL with Q-network and Target net where the LSTM network layer is used in Q-network and the candidate network is used to update Target Net.
DPM framework \citep{b177} based on RL, adopts the long short-term memory (LSTM) network to capture the prediction results and uses DRL to train the strategy of resource allocation aimed at reducing energy consumption in the Cloud environment. The LSTM network used to predict the state of the environment can be regarded as a mapper of time-varying in Fig. \ref{fig6}.

DDQN (Dueling Deep Q-Network) \citep{b152} contains a set of convolutional neural networks and a fully connected layer to achieve higher efficiency of data processing, lower network cost, and better security of data interaction.

MRLCO \citep{b161}, a Meta Reinforcement Learning-based method, contains a seq2seq neural network to represent the policy.

MADRL \citep{b168}, a novel multi-agent DRL, contains actor-network and critic-network to generate Q value. The actor-network with the two-layer fully connected network is a mapper from state to action, and the critic-network with two fully connected network hidden layers and an output layer with one node is a mapper from state and action to Q-value.

DRL+FL \citep{b190}, based on DDQN, uses Federal Learning to accelerate the training of DRL agents.

MDP\_DT \citep{b174}, a novel full-model-based RL for elastic resource management, employs adaptive state space partitioning.

RLFTWS \citep{n1} designed a heuristic algorithm for the task allocation and execution according to the selected fault-tolerant strategy, as well as developed a DDQN to select the fault-tolerant strategy adaptively for each task under the current environment state, which is not only prediction but also learning in the process of interacting with the environment.

AV-MPO \citep{n2}, on-policy maximum a posteriori policy optimization with gated transformer-XL, used an attention-based DRL algorithm to a Cloud–edge collaboration manufacturing task scheduling. 

Other DRL-based scheduling algorithms include HCDRL \citep{n4}, DT (decision transformer) using GPT \citep{n5}, CORA \citep{n6}, DRAW \citep{n7}, PRLCC \citep{n8}, ReCARL \citep{n9}, etc. These algorithms still belong to the DRL architecture of Fig. \ref{fig6} analyzed in {Sect.} \ref{sec5}. 

Based on the review and collection of literature, Table \ref{table7} provides a summary of multi-aspects including category and objectives of RL-based algorithms, Table \ref{tableadd7} provides a summary of the mappers, Table \ref{table8} provides the summary of scenario and task/server nature, as well as Table \ref{table9} provides the summary of experimental data and compared baselines. 

\begin{table*}[ht!]
\caption{Summary of RL-based Algorithms in terms of Category and Objectives{.}}
\label{table7}
\setlength{\tabcolsep}{1pt}
\begin{tabular}{p{80pt}p{50pt}p{170pt}}

%\begin{longtable}%最多380
\hline
Algorithm   & Category    & Objectives   \\
\hline
QEEC \cite{b58}        & QL         & Response time, CPU utilization \\
PCRA \cite{n13}& QL & Cost, Qos\\
MDP\_DT \cite{b174}&QL& Cost\\
AGH+QL \cite{b170} & QL& Energy consumption, QoS \\

MRLCO  \cite{b161}&MRL& Network traffic, service latency\\

DeepRM\_Plus \cite{b107} & DRL  & Turnaround time, cycling time \\

DERP \cite{b149}&DRL& Automatic elasticity \\
DPM  \cite{b177}&DRL& Task latency,  energy consumption \\

DQST \cite{b83}& DQL& Makespan, load balancing \\

MDRL \cite{b153}&DDQN& Energy consumption, response time \\
RLTS  \cite{b164}&DDQN& Makespan \\
DRL-Cloud  \cite{b189}&DDQN& Energy consumption, cost \\
ADRL \cite{b169}&DDQN& Resource utilization, response time \\
IDRQN \cite{b151}&DDQN& Energy consumption,  service latency \\
MADRL \cite{b168}&DDQN& Computation delay, channel utilization   \\
DDQN \cite{b152}&DDQN& Service latency, system rewards\\
DRL+FL \cite{b190}&DDQN& Energy consumption, load balancing \\
RLFTWS \cite{n1}& DDQN & Makespan, resource utilization\\
AV-MPO \cite{n2} & DDQN & Customer satisfaction, load balancing\\
\hline
\end{tabular}
\end{table*}

\begin{table*}[ht!]
\caption{Summary of RL-based Algorithms in terms of the Mappers of Decision and Other Mappers.}
\label{tableadd7}
\setlength{\tabcolsep}{2pt}
\begin{tabular}{p{70pt}p{100pt}p{190pt}}

%\begin{longtable}%最多380
\hline
Algorithm   & Mappers of Decision    & Other Mappers\\
\hline
QEEC \cite{b58}        & Q-value table         & - \\
PCRA \cite{n13}& Q-value table & -\\
MDP\_DT \cite{b174}& Q-value table& - \\
AGH+QL \cite{b170} & Q-value table& - \\

MRLCO  \cite{b161}&Seq2seq neural network& -\\

DeepRM\_Plus \cite{b107} & CNN of six layers  &  - \\
DERP \cite{b149}&FC& - \\
DPM  \cite{b177}& FC&  LSTM  to capture the
prediction results \\

DQST \cite{b83}& FC& - \\

MDRL \cite{b153}&Action-value Q network& Using target-Q network as mapper of feedback \\
RLTS  \cite{b164}&Action-value Q network& Using target-Q network as mapper of feedback \\
DRL-Cloud  \cite{b189}&Action-value Q network& Using target-Q network as mapper of feedback \\
ADRL \cite{b169}&Action-value Q network& Using target-Q network as mapper of feedback.  Neural network to perceive the state of environment \\
IDRQN \cite{b151}&LSTM & Using target-Q network as mapper of feedback \\
MADRL \cite{b168}&Actor-critic network& Using target network as mapper of feedback  \\
DDQN \cite{b152}&A set of CNN and FC & Using target-Q network as mapper of feedback \\
DRL+FL \cite{b190}&Two FC of DNNs& Using target network as mapper of feedback  \\
RLFTWS \cite{n1}& FC & Using target-Q network as mapper of feedback\\
AV-MPO \cite{n2} &  Attention network & Using target network as mapper of feedback\\
\hline
\end{tabular}
\end{table*}

\begin{table*}
\caption{Summary of RL-based Algorithms in terms of Scenario  and Task/Server Nature.}
\label{table8}
\setlength{\tabcolsep}{2pt}
\begin{tabular}[t]{p{65pt}p{145pt}p{150pt}}

%\begin{longtable}%最多380
\hline
Algorithm   & Scenario & Task/Server Nature    \\
\hline
QEEC \cite{b58}        & Online task scheduling        & Independent heterogeneous servers \\
PCRA \cite{n13}& Dynamic resource scheduling & Independent tasks\\
MDP\_DT \cite{b174}&Dynamic resource scheduling& Dynamic tasks\\
AGH+QL \cite{b170} & Resource scheduling in C-RANs& Traffic demand in wireless networks\\

MRLCO \cite{b161}&Adaptive task offloading& Multiple tasks with inner dependencies\\

DeepRM\_Plus \cite{b107} & Online resources scheduling & Independent tasks \\
DERP \cite{b149}&Dynamic resource scheduling& Dynamic tasks \\
DPM \cite{b177}&Online resources scheduling& Dynamic tasks\\

DQST \cite{b83}& Dynamic online task scheduling& Non-preemptive task, heterogeneous server\\

MDRL \cite{b153}&Dynamic resource scheduling& Depended tasks, heterogeneous servers \\
RLTS \cite{b164}&Dynamical tasks scheduling& Depended tasks, heterogeneous servers\\
DRL-Cloud \cite{b189}& Resource provisioning& Tasks with dependencies\\
ADRL \cite{b169}&On-time VMs  scheduling& Dynamic tasks\\
IDRQN \cite{b151}&Task offloading& Depended Tasks; heterogeneous servers \\
MADRL \cite{b168}&Multichannel access and task offloading& Joint multichannel access\\
DDQN \cite{b152}& Online resource scheduling& Delay-tolerant data computing tasks\\
DRL+FL \cite{b190}&Dynamic resource scheduling& Dynamic tasks\\
RLFTWS \cite{n1}&Dynamic workflow scheduling& Dynamic dependent tasks; heterogeneous server\\
AV-MPO \cite{n2} &Dynamic  task scheduling & Dynamic dependent tasks\\
\hline

\end{tabular}
\end{table*}

\begin{table*}[ht!]
\caption{Summary of RL-based Algorithms in terms of Experimental Data  and Compared Baselines.}
\label{table9}
\setlength{\tabcolsep}{2pt}
\begin{tabular}[t]{p{65pt}p{130pt}p{160pt}}

%\begin{longtable}%最多380
\hline
Algorithm   & Experimental Data  & Compared Baselines \\
\hline
QEEC \cite{b58}        & Simulated data by CloudSim& MMS-RANDOM, MMS-FAIR, MMS-GREEDY, basic QL and improved QL\\
PCRA \cite{n13}& Simulated data & ML-based and rule-based methods\\
MDP\_DT \cite{b174}& Simulated and real data& MDP, QDT, Q-learning\\
AGH+QL \cite{b170} & Simulated data& Pure Q-learning, DRL\\

MRLCO \cite{b161}&Simulated data &Fine-tuning DRL, HEFT-based, Greedy\\

DeepRM\_Plus \cite{b107} & Simulated data, Alibaba-Cluster-trace-v2018& Random, FCFS, SJF, HRRN, Tetris, DeepRM\\
DERP \cite{b149}& Okeanos service' data& MDP, Q-learning, MDDPT, QDT\\
DPM \cite{b177}&Google cluster traces& DRL, Round-robin \\

DQST \cite{b83}& Simulated data by WorkflowSim& FCFS, MAXMIN, MCT, MINMIN, RR\\

MDRL \cite{b153}&Simulated data by CloudSim& FIFO, Greedy algorithm\\
RLTS \cite{b164}&Simulated data& HEFT, CPOP, Lookahead, PEFT\\
DRL-Cloud \cite{b189}&  Google cluster-usage traces& Greedy, FERPTS, Round-robin\\
ADRL \cite{b169}&Simulated data by CloudSim& Over-utilized, Under-utilized, DRL\\
IDRQN \cite{b151}& Simulated data by iFogSim& DQN, HERDQN, IDQN, DRQN\\
MADRL \cite{b168}&Simulated data by TensorFlow& Actor-critic; DQN; Greedy\\
DDQN \cite{b152}& Simulated data &Conventional DQN, Greedy, Random\\
DRL+FL \cite{b190}&Simulated data &Centralized DDQN, DRLRA, SDR, LOBO\\
RLFTWS \cite{n1}& Simulated data & RPFTWS, RSFTWS, RI, NC\\
AV-MPO \cite{n2} &  Real data from CMfg& PPO, SAC, DDQN\\
\hline

\end{tabular}
\end{table*}

\begin{table*}[ht!]
\caption{Summary of RL-based Algorithms in terms of Strategies and Advantages.} 
\label{table10}
\begin{small}

\setlength{\tabcolsep}{2pt}
\begin{tabular}[ht!]{p{75pt}p{280pt}}

%\begin{longtable}%最多380
\hline
Algorithm   & Strategies and Advantages\\
\hline
QEEC \cite{b58}        & M/M/S to reduce the average waiting time of task; dynamic task ordering strategy to promote the quality of Cloud services\\
PCRA \cite{n13}&  Multiple prediction learners for making accurate Q-value prediction, which make automatic decisions through interacting with the environment without prior knowledge\\
MDP\_DT \cite{b174}& Adaptively partitions the state space utilizing novel statistical criteria and strategies to perform accurate splits\\
AGH+QL \cite{b170} & Anchor graph hashing can accelerate training; hash codes can reduce size of state space\\

MRLCO  \cite{b161}&Seq2seq neural network to represent the offloading policy; new training method combining the first-order approximation\\

DeepRM\_Plus \cite{b107} & Imitating learning to accelerate convergence and CNN to capture the state of resource\\

DERP \cite{b149}&DERP does not demand space Partitioning; DERP with three aspects manages to collect rewards\\
DPM \cite{b177}& Using LSTM Network to predict workload which can eliminate the vanishing gradient problem\\

DQST \cite{b83}& Entropy weight method to produce a high-quality solution of bi-objective optimization\\

MDRL \cite{b153}&DRL can adapt to scalable state space; fair resource allocation helps reduce the underlying practical problems\\
RLTS  \cite{b164}&Utilization of DQN to describe the relationship between state-agent and action\\
DRL-Cloud  \cite{b189}&  Experience replay, target networks as well as exploration and exploitation can accelerate converge speed\\
ADRL \cite{b169}&Using an anomaly detection model to identify performance problems and to increase awareness of  the environment\\
IDRQN \cite{b151}& LSTM to estimated value; candidate networks to decouple the action selection and action value evaluation\\
MADRL \cite{b168}&Combination of actor-critic and DQN can improve performance of algorithm\\
DDQN \cite{b152}& DDQN can keep stability of reward\\
DRL+FL \cite{b190}&Combination of DRL and FL can improve the performance in training\\
RLFTWS \cite{n1}& Two groups of neural networks with the same structure are designed to reflect the two different goals\\
AV-MPO \cite{n2} & The transformer layers can better perceive scheduling status and evaluate the quality of optimized solutions\\
 
\hline
\end{tabular}

\end{small}
\end{table*}

From Table \ref{table7}, the DRL method mainly using QL and DQN can address a variety of optimization objectives almost covering the existing optimization objectives.
Based on the analysis for the evolution of the DRL framework in {Sect.} \ref{sec5}, we can generalize these algorithms from the perspective of mappers. In Table \ref{tableadd7}, DNNs are mainly used to be the decision-makers of DRL in scheduling. In DDQN, a target network is used as the mapper of feedback.  This is consistent with our analysis in {Sect.} \ref{sec4} and {Sect.} \ref{sec5} that the two key factors for scheduling are production and evaluation of schemes. This also indicates that these two factors are the difficulties in scheduling problems. By incorporating different strategies and networks into the corresponding mapper in Fig. \ref{fig6}, we can obtain the corresponding RL-based scheduling algorithms. On this basis, the additional strategies are mainly aimed at improving training speed, perception ability, accuracy of the evaluation for the solution and optimality. 
From Table \ref{table8}, DRL methods are mainly used for dynamic or online scheduling, and they are not only applicable to heterogeneous resources and independent tasks but also to dependent tasks and non-preemptive tasks, which demonstrates DRL methods have wider application scenarios. From Table \ref{table9}, DRL methods outperform many existing scheduling algorithms. This re-verifies our analysis in {Sect.} \ref{sec4} that before actually executing the tasks in server nodes, classic algorithms cannot accurately evaluate or predict the quality of optimization schemes in dynamic scheduling of complex scenarios. Using DNN, DRL can obtain the performance of a scheme and guide for further improvement of the solution.

\begin{table*}
\caption{Future Work of RL-based Algorithms.}
\label{table11}
\setlength{\tabcolsep}{3pt}
\begin{tabular}[ht!]{p{70pt}p{280pt}}
%\begin{longtable}
\hline
Algorithm   & Future Work\\
\hline
QEEC \cite{b58}        & To investigate meta-heuristic to increase the performance; to establish various queuing model to satisfy realistic scenes\\
MDP\_DT \cite{b174}& To combine the strategies of MDP\_DT with DQN to solve complex scenarios\\
AGH+QL \cite{b170} & To combine anchor graph hashing with DRL\\

MRLCO  \cite{b161}&To apply an adaptive client selection algorithm to automatically filter out stragglers \\

DeepRM\_Plus \cite{b107} &  To apply other policies such as Actor-Critic network and DDPG; to analyze the state  recognition analytically\\

DERP \cite{b149}&To combine DERP with federate learning; to design intercommunicate framework of simple DRL, full DRL and DDRL\\
DPM  \cite{b177}&To combine LSTM predictor and CNN  to reduce energy\\

DQST \cite{b83}& To establish model of energy consumption or multi-objective\\

MDRL \cite{b153}&To consider dependent tasks and workflow\\
DRL-Cloud  \cite{b189}&  To utilize it in dynamic scheduling and static scheduling\\
ADRL \cite{b169}& To applied  parameter initialization strategy and combination of DRL and semi-supervised learning  to accelerate training\\
IDRQN \cite{b151}& To apply transfer learning to heterogeneous MEC ; to utilize federate learning to solve multi-objectives problems\\
MADRL \cite{b168}&To use gated recurrent units of the network to predict channel conditions\\
DDQN \cite{b152}& To apply it in other issues such as energy efficiency\\
DRL+FL \cite{b190}&To utilize the combination of FL and DRL in other scenarios\\
RLFTWS \cite{n1}& More objective and the deadline constraints of workflows can be considered \\
AV-MPO \cite{n2} & More attention is paid to the improvement of the training efficiency of existing algorithms. More attention-based RL algorithms will be applied to solve multi-objective problems\\
\hline
\end{tabular}
\end{table*}

In {the} reviewed literature, strategies of queuing, accelerating training,  partitioning state space of the agent, capturing resource state, keeping the stability of rewards, etc., are proposed to optimize the performance of algorithms. In order to more accurately analyze the advantages of the DRL methods (or RL) in this literature, we collect the advantages of each literature according to the description in the corresponding literature. Then, their details are listed in Table \ref{table10}.
Combined with the results and conclusion in the reviewed literature, future work of DRL-based algorithms (or RL) in reviewed literature are listed in Table \ref{table11}. With the structural information listed in tables, we can deeply discuss existing DRL-based methods in Cloud scheduling.

\begin{figure}[ht!]
\centering
        \begin{minipage}[h]{0.98\columnwidth}
            \centering
            \includegraphics[width=\columnwidth]{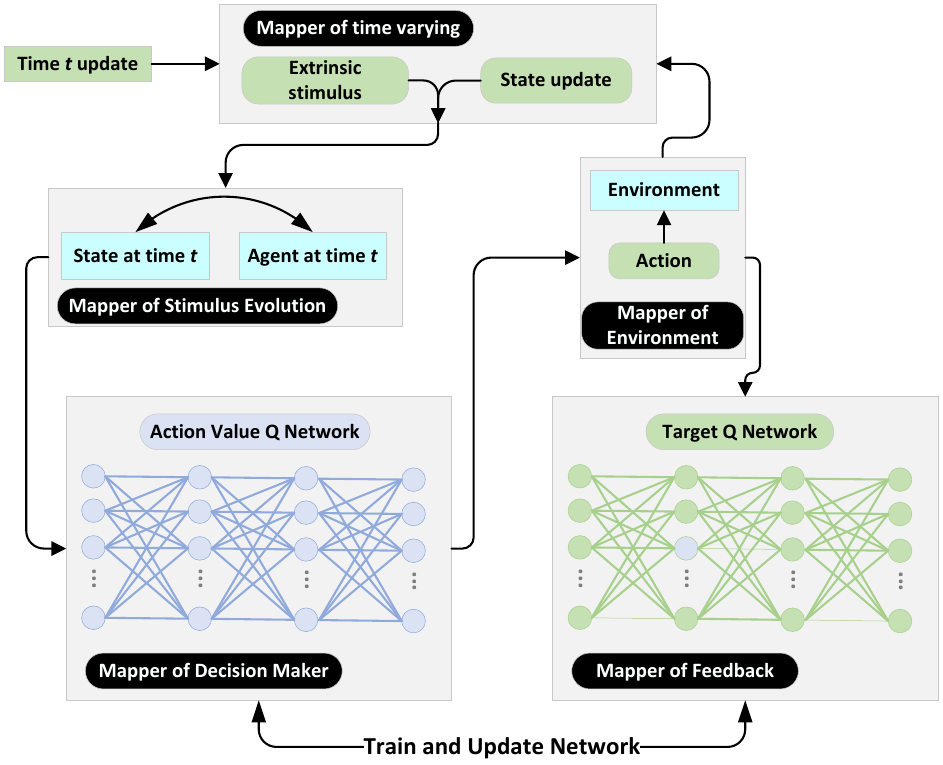}
            \centering
            \caption{A Framework of DDQN of DRL for Scheduling.}
            \label{fig8}
        \end{minipage}
        \end{figure}

\subsection{Discussion}
Based on the above review of RL-based Cloud scheduling especially based on the information listed in Table \ref{table10} and Table \ref{table11},
we summarize the current situation and advantages of DRL (and RL) in Cloud scheduling as follows.
\begin{itemize}[leftmargin=*, topsep = 0em, partopsep=0em, itemsep=0em, parsep = 0em]
\item DRL has strong adaptability for continuous or high dimensional state space; adaptability for scheduling scenarios and various optimization objectives of Cloud computing.
\item DRL has the flexibility to adopt various DNNs as the mappers to predict some implicit information, so as to improve the optimality of scheduling.
\item The main scenario closest to a realistic scene used RL to solve in reviewed literature is a dynamic online multi-resources scheduling problem in a Cloud computing environment or Edge-Cloud computing environment which can contain dependent or independent tasks, workflows, and homogeneous or heterogeneous servers. 
\item In the reviewed literature, experiment results showed DRL can achieve better performance than various commonly compared algorithms such as Randomization, FCFS, Round-robin, Greedy, Q learning, MDP, QDT, FIFO, HEFT, FA, and SDR. And these algorithms together with Conventional DQN can be regarded as baselines to evaluate other algorithms in the future.
\item DDQN is the most commonly used model to solve the scheduling problem in Cloud computing as well as in some Edge-Cloud computing of reviewed literature \citep{b153, b164, b189, b151, b176, b169, b152, b168, b190}. Its common framework can be drawn as Fig. \ref{fig8}. The DDQN contains two networks, action-value Q network and target-Q network, with the same structure \citep{b190,b200}. The action-value Q network can generate the Q-value of the action corresponding {to the} current state. Additionally, the target-Q network can generate target value based on real-time feedback and long-term feedback to obtain the loss function to train the action-value Q network. Simultaneously, the wide application of DDQN also shows that the DRL model has strong adaptability and portability for various scheduling scenarios and optimization objectives.
\item DRL can also be leveraged to address multi-objective scheduling problems, while the previous methods to solve multi-objective optimization problems are mainly meta-heuristic algorithms.
\item The major policies in reviewed literature using DRL (or RL) contain several aspects:
    \begin{itemize}[leftmargin=*, topsep = 0em, partopsep=0em, itemsep=0em, parsep = 0em]
    \item adjusting the structure of decision-mapper to DNN or Q-table;
    \item  strategies to accelerate training of (deep) RL such as the periodical update;
    \item  partition strategies for state-space;
    \item  federal learning to improve convergence and stability;
    \item  strategies to perceive current states or to predict subsequent states of the agent in RL;
    \item  policies to provide loss function to train main-net in DRL.
    \end{itemize}
\end{itemize}

\section{Challenges and Future Directions for DRL-based Scheduling}\label{sec7}
With the comprehensive review and analysis of the previous sections, we can discuss the challenges and future direction of DRL in Cloud scheduling.

Although DRL-based scheduling algorithms have performed advantages in the reviewed literature. 
DRL, as a complex, non-analytic and time-costing algorithm \citep{b196}, has inevitable challenges to address the scheduling problems in real large-scale Cloud computing systems. Based on the comprehensive review of the existing Cloud scheduling and the investigation of the actual operation process, we collect the main challenges and defects using DRL to solve scheduling problems in Cloud computing as follows:
\begin{enumerate}[label=(\arabic*), leftmargin=*, topsep = 0em, partopsep=0em, itemsep=0em, parsep = 0em]%[itemindent=-2em]
    \item	DRL consumes large computing power and occupies prodigious complexity in the progress of training and computation especially for multi-clusters or large-scale systems. Thus, DRL requires a certain period of time before it can be put into use. For scenarios with single and computable objectives, using DRL is not cost-effective.
    \item	The scheduling results based on DRL are still unpredictable, so the performance of the worst case is hard to evaluate. Therefore, the probability of system collapse caused by extremely poor scheduling schemes is not 0.
    \item	Real scheduling also depends on the prediction of dynamic tasks without preemptive and prior knowledge. Since DRL is based on DNN to perceive system states and generate solutions, its performance is highly dependent on the accuracy of DNNs. The training dataset of DRL is limited to cover scenarios in the real system. Thus, it is necessary to retrain the DRL model for a new scenario.
    \item	Gradient descent algorithm used in DRL or Bellman Equation used in QL have inherent restrictions which will lead to local optimization rather than global optimization. Additionally, the training labels of DRL in scheduling are usually not the global optimal solutions. There is currently a lack of public reliable datasets for its training. These will result in a significant gap between the DRL solution and the theoretically optimal solution.
    \item	Unexplainability of training process challenges for the theoretical derivation based on mathematics techniques. Modeling and theoretical derivation of high dimensional continuous state space demand the development of mathematics.
    \item	Antagonism network, federal learning, mathematical logic, Nonhomogeneous Markov process, hidden Markov process and other policies can be utilized in DRL or RL while existing DRLs are mainly based on homogeneous Markov process. For Nonhomogeneous Markov processes, the results involve the solution of matrix differential equations with variable coefficients, which is difficult currently.
    \end{enumerate}

The solution to these challenges requires more pursuit of theoretical research based on the improvement of mathematical theory such as Markov Decision Processes \citep{b196,b170}, Gradient Descent Theory \citep{b200,b107}, Matrix theory \citep{b170} and Discrete Mathematics,  as well as requires more research of real objects in realistic scenarios such as thermal conversion process, calculation process, driving process by electric signal and voltage switching process over the time of components.
Nevertheless, research on DRL utilizing in Cloud scheduling still has considerable potential, which can advance modeling for complex scenes, theoretical modeling for ML, reduction of computational complexity, the flexibility of scheduling algorithms and theoretical research on existing algorithms in principle. 

%In addition,  in resource scheduling of Cloud computing, DRL can be used not only as a resource scheduling algorithm directly like reviewed literature in Section \ref{sec6}, but also as an algorithms selector like Figure \ref{fig10} or DRLS in \cite{z1}. 

According to the previous review and analysis, some of the future directions utilizing RL especially DRL in Cloud scheduling can be summarized as follows.

\begin{figure}[ht!]
        
            \centering
            \includegraphics[width=0.8\columnwidth]{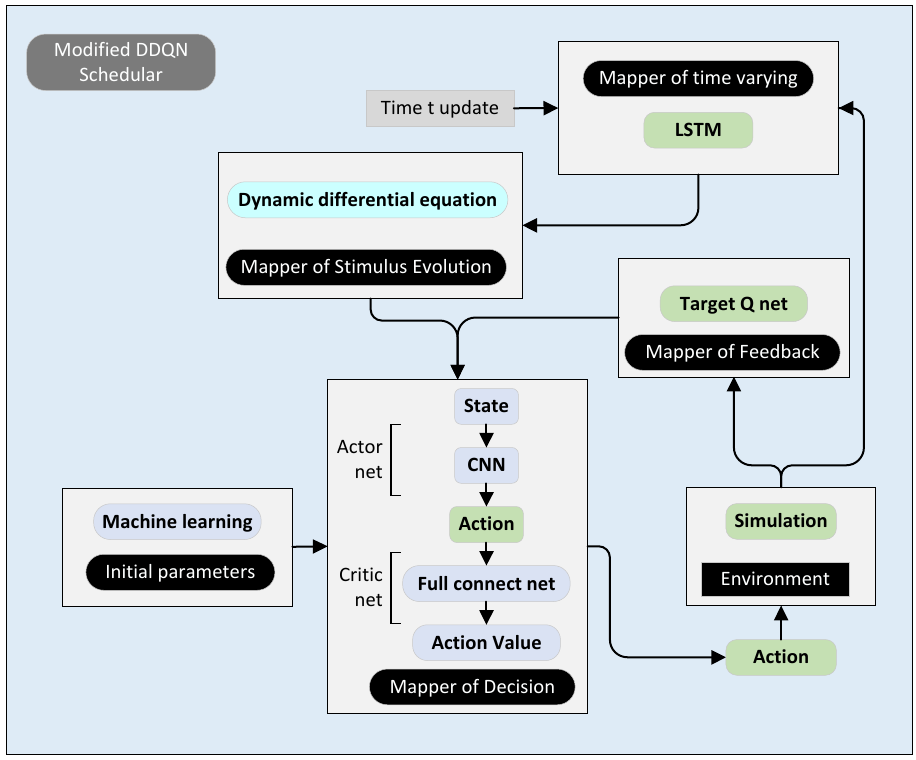}
            \centering
            \caption{Framework of Modified DDQN {C}ombined Various Approaches and Networks of References.}
            
            \label{fig9}
      
   \end{figure}

\begin{enumerate}[label=(\arabic*), leftmargin=*, topsep = 0em, partopsep=0em, itemsep=0em, parsep = 0em]%[itemindent=-1em]
\item	DRL can combine other policies or approaches to meet complex scenes and multi-objectives, which has been verified in experiments of reviewed literature. Therefore, its application mode in a realistic engineering environment to reduce the risk of excessive computational complexity is a noteworthy direction. 
\item	As the RL is one category of non-supervised learning, one of the crucial issues is how to train the main network of the decision-mapper in DRL. The main net in DRL can be represented by CNN, LSTM, Transformers, etc. Setting various training strategies can accelerate the convergence speed of DRL, enabling it to participate in generating scheduling schemes more quickly.
    Some policies to improve the convergence performance of DRL contain leverages of meta-heuristic method and imitating learning \citep{b107}. Queue model is also a crucial aspect to increase the performance of schedulers such as M/M/S queuing model \citep{b58} and M/G/1 queuing model \citep{b45}.
\item	Application pattern of DRL to assist other analyzable scheduling algorithms such as FCFS, Hungarian algorithm, LPT algorithm, Johnson's algorithm and more.

%\item	Another emergency trend of resource scheduling in Cloud computing is the application of algorithms in real complex and random scenarios.
\item	It a potential direction to adjust and improve the mappers based on the existing architecture of Fig. \ref{fig6}, so as to widen the adapted scenarios and improve the optimality of DRL.
    E.g., combining with the reviewed literature and our research, we integrate and gain a framework of modified DDQN combined various approaches and networks to solve scheduling problems of Cloud computing, especially for realistic scenarios as Fig. \ref{fig9}. 
\item	The exploration of more novel roles of DRL is also a practical research direction, e.g., DRL can perform as a system strategist to boost the process of selecting methods, as specific approaches are able to adapt specific scenarios \citep{z1}. To illustrate its possible novel role, a Deep Q-learning-based framework of the scheduler is presented in Fig. \ref{fig10} used to schedule various scheduling algorithms for different scenarios, which can be called a scheduler of scheduling algorithm aiming to give full play to the superiorities of all scheduling algorithms and considering that all the algorithms are part of anthroposophy. In this framework, the scheduling algorithms are regarded as resources that can be automatically selected and DRL-based algorithms are not only components of resource scheduling algorithms but also strategies to guide the selection of specific scheduling algorithms.  DRL-based algorithm selector is one such attempt, whose framework can be seen in Fig. \ref{figadd6} \citep{z1}. From \cite{z1}, it is not recommended to use a single algorithm for all scenarios.

\begin{figure}[ht!]
\centering
        
            \includegraphics[width=\columnwidth]{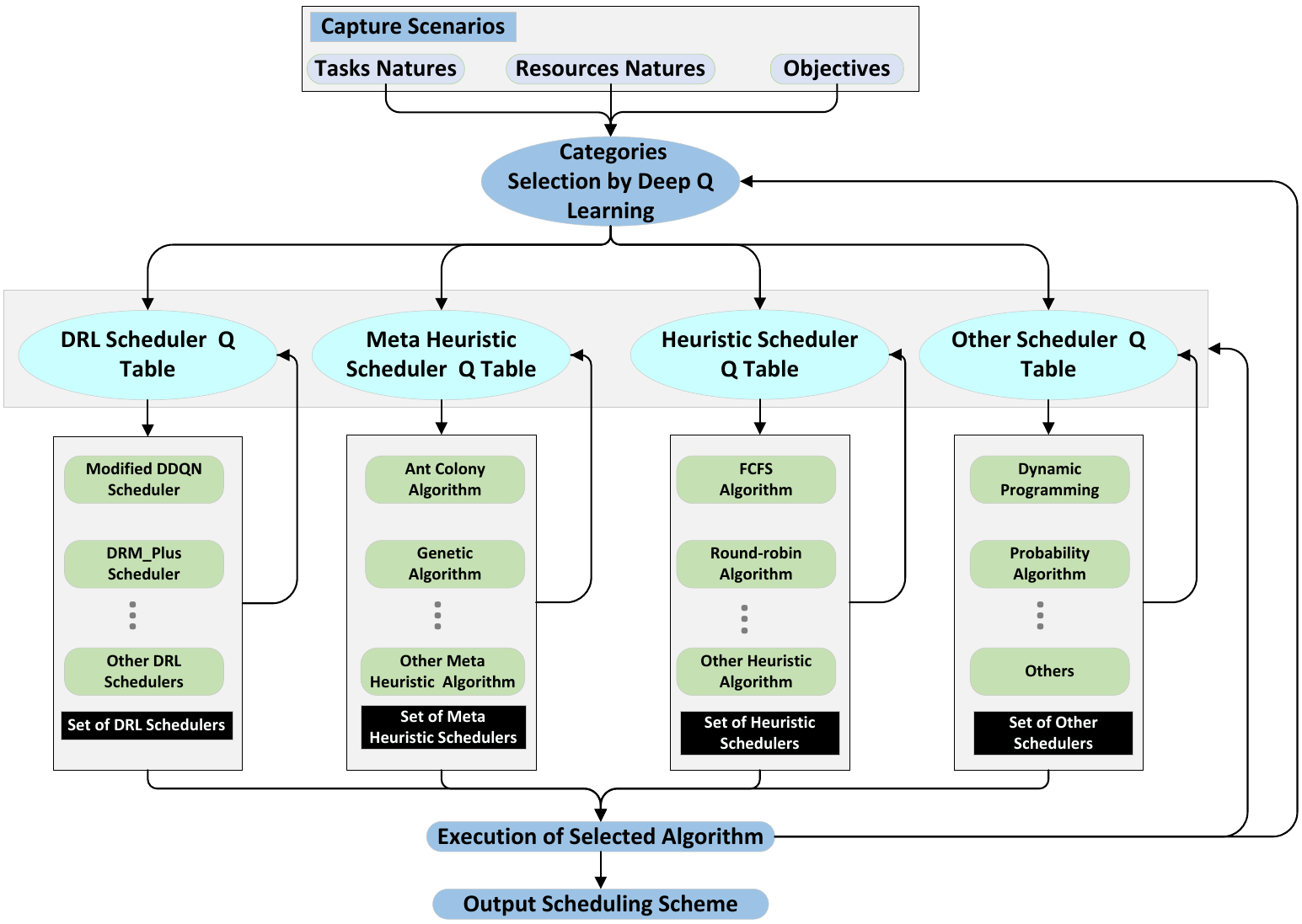}
            \caption{Two Phases Q Learning-based Scheduler used to Schedule Various Scheduling Algorithms.}
            \label{fig10}
       
\end{figure}

\begin{figure}[ht!]
            \centering
            \includegraphics[width=0.99\columnwidth]{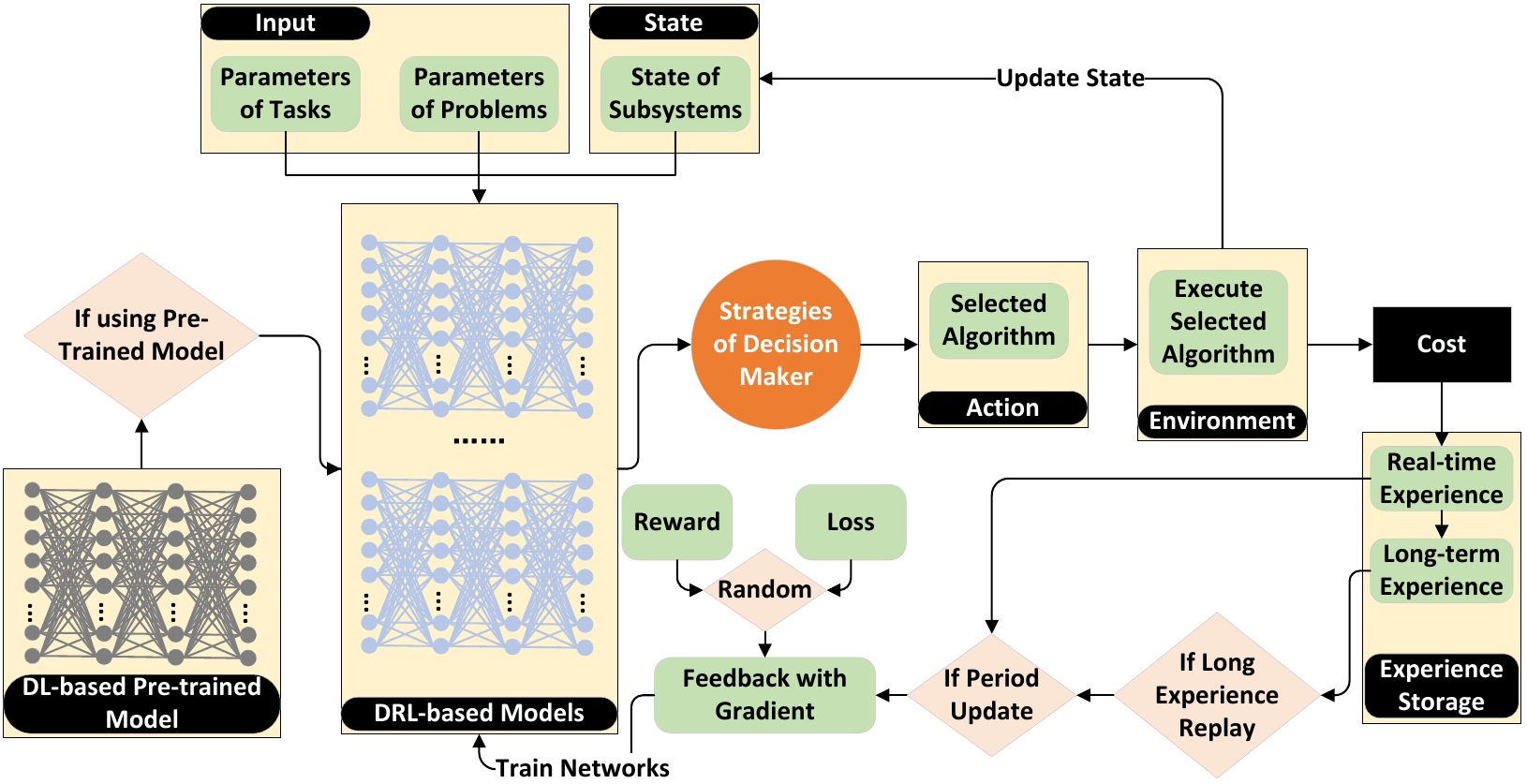}
            \caption{A Framework of DRL-based Selector with Various Strategies \citep{z1}.}
            \label{figadd6}
\end{figure}    

\item	Additional bases for the application of DRL in scheduling problems are baselines and benchmarks. We summarize the characteristics of baselines and benchmarks as follows.

\textbf{A baseline should possess the following properties.}
    \begin{enumerate}[itemindent=1em, leftmargin=0em, topsep = 0em, partopsep=0em, itemsep=0em, parsep = 0em]
    \item	The baseline is easy to construct and implement;
    \item	It has reproducibility and performance stability;
    \item	It can adapt to multiple scenes, objectives and experiments with variable scales;
    \item	It should include various types of algorithms;
    \item	Its performance can be proved theoretically or has recognized conclusion;
    \item	It should have been optimized to a certain extent and should contain; some state-of-the-art representing the characteristics of their types;
    \item	The compared experiments should reduce the influence of parameter adjustment as much as possible and reserve the inherent performance of the algorithm.
    
    \end{enumerate}
\textbf{A benchmark should possess the following properties.}
    \begin{enumerate}[itemindent=1em, leftmargin=0em, topsep = 0em, partopsep=0em, itemsep=0em, parsep = 0em]
    \item	It should be easy to reproduce and calculate;
    \item	It should contain data from multiple scenarios and be as close as possible to real scenarios;
    \item	It can be applied to the experimental verification of a variety of optimization objectives;
    \item	It should have a dynamic scale rather than a single scale which can avoid the performance optimization caused by adjusting parameters;
    \item	If there is a random experiment, a benchmark should have enough sampling times;
    \item	It should have certain control variables to verify the advantage of local strategy;
    \item	It should contain enough extreme scenarios, especially on some parameter boundaries;
    \item   The comparison is relatively fair, such as comparing the solutions generated under the same computational cost;
    \item	It can test the algorithm running under a variety of devices and components.
    \end{enumerate}

\item Other potential directions used DRL to solve scheduling problems in Cloud computing still demand further research, which can be listed as follows.
      \begin{enumerate}[itemindent=1em, leftmargin=0em, topsep = 0em, partopsep=0em, itemsep=0em, parsep = 0em]
      \item How to construct a novel well-performed framework of RL or DRL and how to construct a novel well-performed DNN in DRL?
      \item How to accelerate the training or reduce the calculation complexity of (deep) RL to enhance its transferability?
      \item How to ensure the stability of the results under the application of DRL to resolve scheduling problems in large-scale Cloud computing to avoid the risk caused by extremely poor schemes?
      \item How to construct the deducible optimization theory?
      %\item How to predict the running state of Cloud systems?
      \item How to build a flexible scheduling system combining various scheduling algorithms to cope with time-varying objectives?
      \item How to capture agent-state of DRL accurately?
      %\item How to construct the communication between multi-DRL agents of federal learning?
      \item How to improve other categories of methods to address pervasive scheduling problems not only in Cloud computing but also in other distributed systems?
      \end{enumerate}
\end{enumerate}

\section{Conclusions}\label{sec8}
In this paper, we provide a universal formulation of scheduling and review various types of scheduling algorithms in Cloud. Two key factors of scheduling are the production and evaluation of solutions.
By analyzing the formulation and algorithms of scheduling, we discuss the defects of classic algorithms, which also demonstrate the necessity of DRL-based methods for scheduling. To assist the acquaintance of DRL in Cloud scheduling, we provide the analysis for the evolution of RL frameworks (including DRL) from the perspective of mappers. On the basis of analysis for RL frameworks, we provide a survey of existing DRL-based methods in Cloud scheduling. Then, we analyze and discuss the advantages, challenges and future direction of {DRL-based Cloud scheduling}.

From this surveyed work, we can see that the application of DRL in resource scheduling of Cloud computing is an effective and non-substitutable technique. Simultaneously based on the reviewed literature, some of the main advantages of DRL used in resource scheduling are adaptability and portability to scenarios and optimization objectives because the use of DNN enables DRL to describe the higher dimensional and or continuous agent's state space when the objective is implicit or hard to calculate, which allows DRL-based methods to achieve better performance in many complex scenarios such as the dynamic resource scheduling of large-scale Cloud computing for dependent tasks and heterogeneous servers. Due to the combination of DL and RL, DRL-based scheduling algorithms can solve some scheduling problems that the classic algorithms are unable to solve.

With the reviews of existing works, we discuss the challenges of DRL in Cloud scheduling.  The main challenges of using DRL in Cloud scheduling are complexity, unexplainability and local convergence of the training process, as well as the unpredictability of scheduling results. Then, we provide several potential directions for future research of DRL in Cloud scheduling based on these challenges. In future directions, in addition to combining other policies to reduce the complexity and improving structures of DRL, we also propose a point of view of using DRL as an algorithm selector for scheduling. Moreover, we also list the properties required for {the} baseline and benchmark of DRL-based Cloud scheduling. 

Based on the above work in this paper, we can see that DRL has significant potential in Cloud scheduling deriving abundant research directions.  Regarding all algorithms as resources, how to combine DRL with other types of algorithms to solve more difficult scheduling problems (i.e., scheduling algorithm selectors) is still worthy of continuous exploration and research.

\section*{Acknowledgments}
This research is partially supported by the National Key Research and Development Program of China with ID 2018AAA0103203.2, by the Project of Key Research and Development Program of Sichuan Province with Grant ID 2021YFG0325, and by the Sichuan Provincial Science and Technology Plan Project with Grant ID 2021JDRC0005.
%\bibstyle{sn-mathphys}
%\bibstyle{sn-basic}

\bibliographystyle{elsarticle-num}
\bibliography{mybibfile}

\end{sloppypar}

\end{document}